\documentclass{bioinfo}

\usepackage{textcomp}
\usepackage{natbib}
\usepackage{afterpage} 

\usepackage{placeins}
\usepackage{float}
\usepackage{capt-of}

\usepackage[margin=0pt,font=normalfont,labelfont=bf,labelsep=period]{caption} 

\usepackage{balance} 
\usepackage{fancyvrb}
\usepackage{bbm} 
\usepackage{adjustbox}
\usepackage{stackengine,scalerel}

\usepackage{booktabs}
\usepackage{url}
\usepackage[11pt]{moresize}
\renewcommand{\marginpar}[1]{}
 
\usepackage{mathtools}
\usepackage{times}
\usepackage{latexsym}
\usepackage{multirow}
\usepackage{arydshln}
\usepackage{graphicx}
\usepackage{grffile}
\usepackage{ulem}	
\usepackage{pdfpages}
\usepackage{xspace}
\usepackage{amssymb,amsmath,epsfig}
\usepackage{mathpartir}
\usepackage{amsthm}
\usepackage{mathrsfs}
\usepackage{algorithm}
\usepackage{pifont}
\usepackage[noend]{algpseudocode}
\usepackage{tikz}
\usetikzlibrary{matrix, positioning, arrows.meta, calc, shapes, decorations,
  backgrounds, arrows, decorations.pathreplacing} 
\usepackage{pgfplots, siunitx}
\usepackage{hhline}
\renewcommand{\cite}[1]{\citep{#1}}

\newcommand{\panel}[1]{\large \sf {#1}}

\newcommand{\nts}{\ensuremath{\text{\it nt}}\xspace}
\newcommand{\pairs}{\ensuremath{\mathrm{pairs}}\xspace}

\newcommand{\hairpinloops}{\ensuremath{\mathrm{hairpin\_loops}}\xspace}
\newcommand{\singleloops}{\ensuremath{\mathrm{single\_loops}}\xspace}
\newcommand{\externalloops}{\ensuremath{\mathrm{external\_loops}}\xspace}
\newcommand{\multiloops}{\ensuremath{\mathrm{multi\_loops}}\xspace}
\newcommand{\unpaired}{\ensuremath{\mathrm{unpaired}}\xspace}
\newcommand{\wunpaired}{\ensuremath{w_\text{unpaired}}\xspace}
\newcommand{\wcg}{\ensuremath{w_\text{CG}}\xspace}

\newcommand{\wau}{\ensuremath{w_\text{AU}}\xspace}

\newcommand{\wgu}{\ensuremath{w_\text{GU}}\xspace}

\newcommand{\ppv}{\ensuremath{\text{PPV}}\xspace}
\newcommand{\sens}{\ensuremath{\text{Sensitivity}}\xspace}

\newcommand{\tuple}[1]{\ensuremath{\langle {#1} \rangle}}

\newcommand{\notes}[1]{}

 \theoremstyle{definition}
 
\theoremstyle{plain}

\newcommand{\ith}[1]{\ensuremath{i^{{th}}}}

\newcommand{\goesto}{\ensuremath{\rightarrow}\xspace}

\newcount\permx
\newcount\permy
\def\permdot#1#2{
\permx=#1 \advance\permx by-1
\permy=#2 \advance\permy by-1
\psframe[fillcolor=black, fillstyle=solid]
(\permx,\permy)(#1, #2)
}

\newcommand{\argmax}{\operatornamewithlimits{\mathbf{argmax}}}

\newcommand{\x}[1]{\ensuremath{x_{#1}}\xspace}

\newcommand{\score}{\ensuremath{\mathit{sc}}\xspace}

\newcommand{\vecx}{\ensuremath{\mathbf{x}}\xspace}
\newcommand{\vecy}{\ensuremath{\mathbf{y}}\xspace}
\newcommand{\vecw}{\ensuremath{\mathbf{w}}\xspace}

\newcommand{\vecystar}{\ensuremath{\vecy^*}\xspace}
\newcommand{\vecyhat}{\ensuremath{\hat{\vecy}}\xspace}
\newcommand{\vecybar}{\ensuremath{\bar{\vecy}}\xspace}
\newcommand{\valid}{\ensuremath{\mathrm{valid}}\xspace}
\newcommand{\balanced}{\ensuremath{\mathrm{balanced}}\xspace}
\newcommand{\depth}{\ensuremath{\mathrm{depth}}\xspace}

\newcommand{\scorew}{\ensuremath{\mathit{sc}_{\vecw}}\xspace}

\newcommand{\smallnt}[1]{\ensuremath{_{\mbox{\tiny PP}}}\xspace}

\iffalse

\else

\fi

\newcommand{\GEN}{\ensuremath{\mathcal{Y}}\xspace}

\newcommand{\smallurl}[1]{{\scriptsize \url{#1}}}

\newcommand{\nitemt}[3]{\ensuremath{\tuple{{#1},\; {#2}}\!: \tuple{#3}\xspace}}
\newcommand{\nnitem}[4]{\ensuremath{\tuple{{#1},\; {#2},\; {#3}}\!: {#4}\xspace}}

\newcommand{\newmid}{\!\mid\!}

\newcommand{\leftb}{\ensuremath{\text{\tt (}}\xspace}
\newcommand{\rightb}{\ensuremath{\text{\tt )}}\xspace}

\newcommand{\mydot}{\ensuremath{\text{\tt .}}\xspace}

\newcommand{\push}{\ensuremath{\mathsf{push}}\xspace}

\newcommand{\pop}{\ensuremath{\mathsf{pop}}\xspace}

\newcommand{\nskip}{\ensuremath{\mathsf{skip}}\xspace}

\newcommand{\ncombine}{\ensuremath{\mathsf{combine}}\xspace}

\newcommand{\md}{\ensuremath{\text{\tt .}}}

\newcommand{\mla}{\ensuremath{\text{\tt (}}}
\newcommand{\mra}{\ensuremath{\text{\tt )}}}
\newcommand{\bml}{\ensuremath{\text{\tt{\textbf (}}}}

\newcommand{\ml}{\mla}
\newcommand{\mr}{\mra}

\newcommand{\mq}{\ensuremath{\text{\tt ?}}}

\newcommand{\weight}{\ensuremath{\mathbf{w}}\xspace}

\newcommand{\ecoli}{{\em E.~coli}\xspace}

\newcommand{\bsubtilis}{{\em B.~subtilis}\xspace}
\newcommand{\cmirabilis}{{\em C.~mirabilis}\xspace}
\newcommand{\linearfold}{{LinearFold}\xspace}
\newcommand{\linearfoldc}{{LinearFold-C}\xspace}
\newcommand{\linearfoldv}{{LinearFold-V}\xspace}
\newcommand{\contrafold}{{CONTRAfold}\xspace}
\newcommand{\contrafoldmfe}{{CONTRAfold MFE}\xspace}
\newcommand{\viennarna}{{ViennaRNA}\xspace}
\newcommand{\viennarnafold}{{Vienna RNAfold}\xspace}
\newcommand{\rnafold}{{RNAfold}\xspace}

\newcommand{\bm}{\mathbf}

\newcommand{\myurl}[1]{\href{{#1}}{\tt {#1}}} 
\newcommand{\myurlsmall}[1]{\scriptsize \url{#1}} 

\newcommand{\jnext}{\ensuremath{\mathrm{next}}\xspace}

\newcommand{\codeblue}[1]{%
\ensuremath{
  \begingroup\setlength{\fboxsep}{1pt}%
  \colorbox{blue!20}{\hspace*{-4pt} \vphantom{Ay} #1 \hspace*{-2pt}}%
  \endgroup
  }
}

\newcommand{\one}{\ensuremath{\mathbbm{1}}\xspace}

\copyrightyear{2019} \pubyear{2019}

\access{ISMB/ECCB 2019} 
\appnotes{Manuscript Category}

\begin{document}
\firstpage{295} 
\renewcommand\thepage{i\arabic{page}} 

\subtitle{Subject Section}  

\title[LinearFold: Linear-Time RNA Folding]{LinearFold: linear-time approximate RNA folding
  by 5'-to-3' dynamic programming and beam search}
\author[Huang \textit{et~al}.]{Liang Huang$^{\text{\sfb 1,2},\ast}$, 
	He Zhang$^{\text{\sfb 2},\dagger}$, 
	Dezhong Deng$^{\text{\sfb 1},\dagger}$, 
	Kai Zhao$^{\text{\sfb 1},\ddagger}$, 
	Kaibo Liu$^{\text{\sfb 1,2}}$, 
	David A.~Hendrix$^{\text{\sfb 3,1}}$ 
	and David H.~Mathews$^{\text{\sfb 4,5,6}}$}
\address{$^{\text{\sf 1}}$School of Electrical Engineering and Computer Science, Oregon State University, Corvallis, OR 97330, USA, 
$^{\text{\sf 2}}$Baidu Research USA, Sunnyvale, CA 94089, USA, 
$^{\text{\sf 3}}$Department~of Biochemistry~\& Biophysics, Oregon State University, and 
$^{\text{\sf 4}}$Department~of Biochemistry~\& Biophysics,
$^{\text{\sf 5}}$Center for RNA Biology, and
$^{\text{\sf 6}}$Department~of Biostatistics \& Computational Biology, University of Rochester Medical Center, Rochester, NY 48306, USA}

\corresp{$^\ast$To whom correspondence should be addressed.\\
  $^\dagger$The authors wish it to be known that these authors contributed equally (co-second authors).\\
  $^\ddagger$: Present address: Google, Inc, New York, NY 10011, USA.}

\history{Received on XXXXX; revised on XXXXX; accepted on XXXXX} 

\editor{Associate Editor: XXXXXXX} 

\abstract{\textbf{Motivation:}   Predicting the secondary structure of an RNA sequence 
  is useful in many applications.
  Existing 
  algorithms (based on dynamic programming)
  suffer from a major limitation:
  their runtimes scale 
%
  {\em cubically} with the RNA length, 
  and this slowness limits their use 
  in genome-wide applications.
  \\
  \textbf{Results:}
  We present a novel alternative $O(n^3)$-time dynamic programming algorithm
  for RNA folding
  that is amenable to heuristics that 
  make it run in $O(n)$ time and $O(n)$ space, 
  while producing a high-quality approximation to the optimal solution.
  Inspired by incremental parsing for context-free grammars in computational linguistics,
  our alternative dynamic programming algorithm
  scans the sequence in a left-to-right (5'-to-3') direction
  rather than in a bottom-up fashion,
  which allows us to employ the effective beam pruning heuristic.
  Our work, though inexact, is the first RNA folding algorithm to 
  achieve linear runtime (and linear space)  without imposing constraints on the output structure.
  Surprisingly, 
  our approximate search results in even higher overall accuracy 
  on a diverse database of sequences with known structures.
  More interestingly, it leads to significantly more accurate predictions 
  on the longest sequence families in that database (16S and 23S Ribosomal RNAs),
  as well as improved accuracies for long-range base pairs (500+ nucleotides apart),
  both of which are well known to be challenging  for the current models.
  \\
  \textbf{Availability:}
  Our source code is available at
  \url{https://github.com/LinearFold/LinearFold}, 
  and our webserver is at
  \url{http://linearfold.org} (sequence limit: 100,000\nts).
  \\
  \textbf{Contact:} \href{mailto:liang.huang.sh@gmail.com}{liang.huang.sh@gmail.com} 
  \\
\textbf{Supplementary information:} \hyperref[sec:extradefs]{Supplementary data} are available at \textit{Bioinformatics}
online (attached \hyperref[sec:extradefs]{here}).
}

\maketitle

\section{Introduction}



Ribonucleic acid (RNA) 
is involved in numerous cellular processes
 \citep{eddy:2001}.  
The dual nature of RNA
as both a genetic material and functional molecule led to the RNA World
hypothesis, that RNA was the first molecule of life \citep{gilbert:1986},  and
this dual nature has also been utilized to develop \textit{in vitro} methods to evolve
functional sequences \citep{joyce:1994}.  Furthermore, RNA is an important drug target
and agent
 \citep{angelbello+:2018,sazani+:2002,crooke:2004,childs-disney+:2007,gareiss+:2008,castanotto+rossi:2009,palde+:2010}. 
%
%

Predicting the secondary structure  of an RNA sequence, defined as the set of all
canonical base pairs (A--U, G--C, G--U, see Fig.~\ref{fig:overall}A), is an important
and challenging problem
\cite{seetin+mathews:2012,hofacker+lorenz:2014}. 
Knowing the structure reveals 
crucial information about the RNA's function, 
which is useful in many applications ranging from ncRNA detection
\cite{gruber+:2010,washietl+:2012,fu+:2015}
to the design of oligonucleotides for knockdown of message
\cite{lu+mathews:2008,tafer+:2008}.
%
%
Since experimentally determining the structure is expensive and
time comsuming,
and given the overwhelming increase in genomic data (about $10^{21}$ base-pairs per year)
\cite{stephens+:2015},
computational methods have been widely used as an alternative to automatically predict the structure.
Widely used systems such as RNAstructure \cite{mathews+turner:2006}, \viennarnafold \citep{lorenz+:2011}, 
CONTRAfold \citep{do+:2006} and CentroidFold~\cite{sato+:2009}, 
%
all use virtually the same dynamic programming (DP)
algorithm \citep{nussinov+:1978,zuker+stiegler:1981} 
to find the best-scoring (lowest free energy, maximum expected accuracy, or best model score) structure
%
%
 \citep{mathews+turner:2006,washietl+:2012}.  
However, this set of algorithms,
borrowed from computational linguistics \citep{kasami:1965,younger:1967}, 
has a running time
of $O(n^3)$ that 
scales {\em cubically} with the sequence length~$n$,
which is too slow for long RNA sequences \citep{Lange+:2012}.

\iftrue
\begin{figure}[t]
  \includegraphics[width=.49\textwidth]{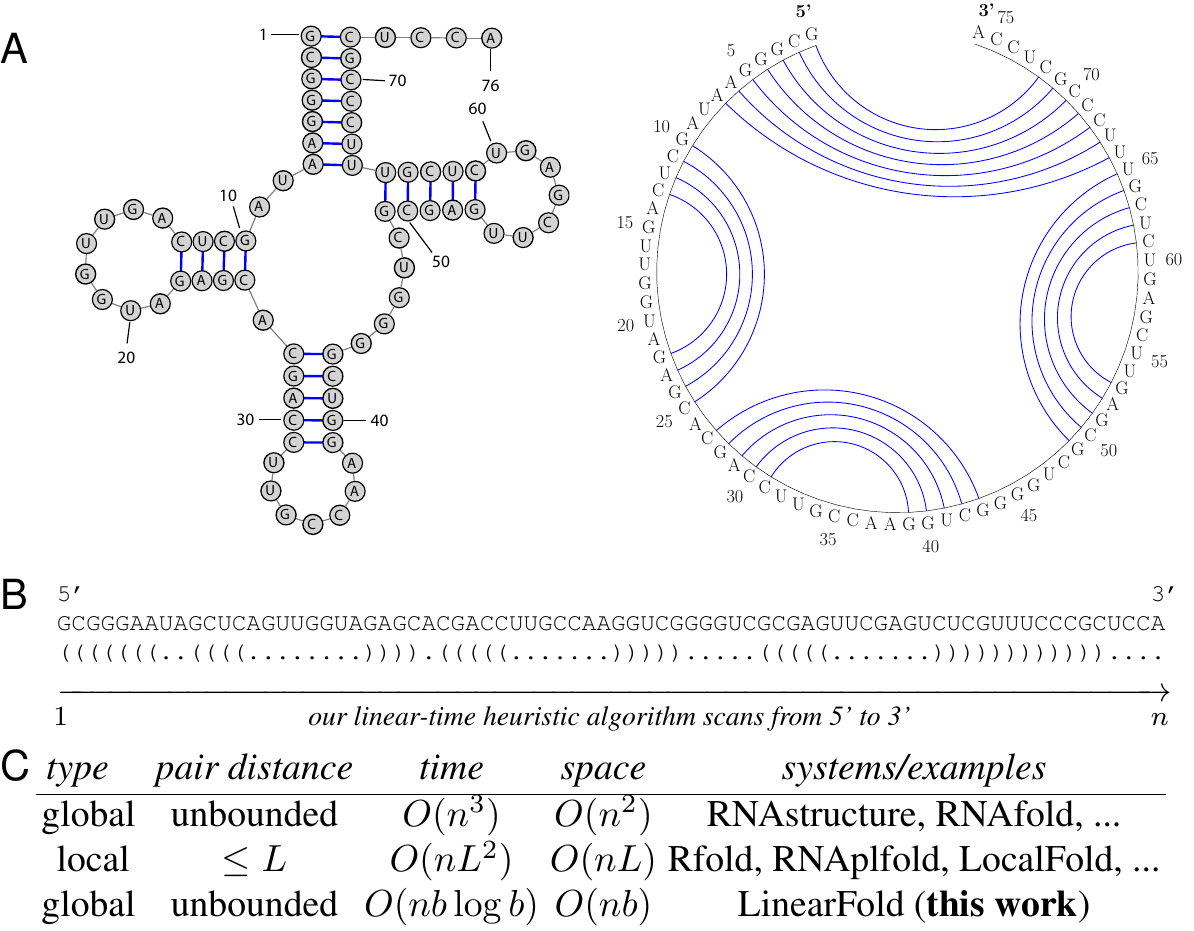}
  \caption{
  Summary of our work. 
    {\bf A}: secondary structure representations of {\it E.~coli} tRNA$^\textit{Gly}$; 
    {\bf B}: the corresponding dot-bracket format and an illustration of our algorithm,
    which scans the sequence left-to-right,
    and tags each nucleotide as ``\md'' (unpaired), ``\ml'' (to be paired with a
    future nucleotide) or ``\mr'' (paired with a previous nucleotide).
    {\bf C}: comparison between our work and existing ones.
    $L$ is the limit of pair distance in local folding methods
    (often $\leq$150), and
    $b$ is the beam size in our work (default 100).
    Our algorithm, though approximate, is the first 
    to achieve linear runtime without imposing constraints on the output structure.
    \label{fig:overall}
  }
  \vspace{-0.4cm}
\end{figure}
\fi

As an alternative,
faster algorithms that predict
only a restricted subset of structures have been proposed.
On the one hand, local folding methods such as Rfold \citep{kiryu+:2008}, Vienna RNAplfold \cite{bernhart+:2006},  and
LocalFold~\cite{Lange+:2012} run in linear time 
but only predict base pairs up to $L$ nucleotides apart 
($L\leq 150$ in the literature; see Fig.~\ref{fig:overall}C).
On the other hand, due to the prohibitive cubic runtime of standard 
 methods,
it has been a common practice to divide long RNA sequences into short segments (e.g., $\leq$ 700\nts)
and predict structures within each segment only
\cite{watts+:2009,andronescu+:2007, licon+:2010}.
All these local methods omit long-range base pairs, 
which theoretical and experimental studies 
have demonstrated to be common in natural RNAs,
especially between the 5' and 3' ends \cite{seetin+mathews:2012,lai+:2018,li+reidys:2018}.
We instead design 
\textbf{LinearFold}, 
an 
approximate  algorithm
that is 
the first in RNA folding 
to achieve linear runtime (and linear space)
without imposing constraints on the output structure
such as base pair distance.
While the classical $O(n^3)$-time algorithm 
is bottom-up, making it hard to linearize, 
ours runs left-to-right (i.e., 5'-to-3'),
incrementally tagging each nucleotide 
in the dot-bracket format (Fig.~\ref{fig:overall}B).
While this naive version runs in the exponential time of $O(3^n)$,
we borrow an efficient packing idea from computational linguistic \citep{tomita:1988} that reduces the runtime back to $O(n^3)$.
This novel left-to-right $O(n^3)$ dynamic program 
is also a contribution of this paper.
Furthermore, on top of this exact algorithm, 
we apply beam search,
a popular heuristic to prune the search space \citep{huang+sagae:2010}, 
which keeps only the top~$b$ highest-scoring (or lowest energy)
states for each prefix of the input sequence,
resulting in an $O(n b \log b)$ time approximate search algorithm,
where $b$ is the beam size chosen by the user. 




Our approach can ``linearize'' any dynamic programming-based pseudoknot-free
RNA folding system. 
In particular,
we demonstrate two versions of LinearFold,
\linearfoldv using the thermodynamic free energy model \citep{mathews+:2004} from Vienna RNAfold \citep{lorenz+:2011}, 
and \linearfoldc using the machine learned model from CONTRAfold \citep{do+:2006}. 
We evaluate our systems on a diverse dataset
of RNA sequences with well-established structures,
and show that while being substantially more efficient,
\linearfold leads to even higher average accuracies
over all families,
and 
more interestingly, 
\linearfold is significantly more accurate than the exact search methods 
on the longest families, 16S and 23S Ribosomal RNAs. 
In addition, 
\linearfold is also more accurate
on long-range base pairs, 
which is well known to be a challenging problem for the current 
models~\cite{amman+:2013}.

Finally, our work
establishes a new connection among
computational linguistics,
compiler theory,
and RNA folding (see Supplementary Fig.~\ref{fig:connect}).

\section{The LinearFold Algorithm}

\vspace{0.2cm}

\subsection{Problem Formulation}
\vspace{0.2cm}

\label{sec:formulation}

Given an RNA sequence
$\vecx = x_1x_2\ldots x_{n}$,  where each $x_i \in \{\tt{A},\tt{C},\tt{G},\tt{U}\}$,
the secondary structure prediction problem
aims to find the best-scoring pseudoknot-free structure \vecyhat
by maximizing a scoring function \scorew (e.g., model score or negative free energy) where $\weight$ are the model parameters:
\vspace{-0.1cm}
\begin{equation}
  \vecyhat = \argmax_{\vecy \in \GEN(\vecx)} \scorew(\vecx, \vecy).
  \label{eq:mfe}
\end{equation}
Here $\GEN(\vecx)$ is the set of all possible pseudoknot-free secondary structures for input \vecx of length $n$
\[
\big\{\vecy \in \{\md, \ml, \mr\}^n \mid \balanced(\vecy), \valid(\vecx,\pairs(\vecy))\big\}
\]
where 
$\balanced(\vecy)$ checks if \vecy has balanced brackets,
$\valid(\vecx,\pairs(\vecy))$ checks if all pairs in \vecy are valid (CG, AU, GU),
and $\pairs(\vecy)$ returns the set of $(i,j)$ pairs where $x_i$ and $x_j$ form a base pair in \vecy,
e.g.,
$\pairs(\text{``\ml\ml\md\mr\mr''}) = \{(1,5), (2,4)\}$.
See Supplementary Section~\ref{sec:extradefs} for detailed definitions.


All dynamic programming-based prediction algorithms, including ours, 
require the scoring function $\scorew(\vecx,\cdot)$ to {\em decompose} to 
smaller structures.
For simplicity of presentation,
in the main text we will use a very simple decomposition to individual pairs and unpaired nucleotides:
\begin{equation}
  \scorew(\vecx, \vecy) = \!\!\!\sum_{(i,j)\in \pairs (\vecy)} \!\!\! w_{x_i x_j} + \!\!\!\sum_{i\in \unpaired(\vecy)} \!\!\! \wunpaired
  \label{eq:simple}
\end{equation}
In this framework we can assign different scores for different pairs, 
and incur a penalty for each unpaired nucleotide. 
For the example in Fig.~\ref{fig:method}, we simply set 
$\wcg\!=\!\wau\!=\!\wgu\!=\!1$
and $\wunpaired\!=\!-0.1$;
therefore, $\scorew(\text{``CCAGG''}, \text{``\ml\ml\md\mr\mr''}) \!=\! 2 \wcg \!+\!\wunpaired \!=\! 1.9$.

In reality, however, the actual scoring functions 
used by \contrafold, \rnafold, and our \linearfold are much more complex,
and they decompose into individual loops.
See Supplementary Section~\ref{sec:realscore} for details.
\newcommand*\circled[1]{\tikz[baseline=(char.base)]{
    \node[shape=circle,draw,inner sep=2pt] (char) {#1};}}

\newcommand{\statescore}[1]{\resizebox{0.3cm}{!}{#1}}

\iftrue
\begin{figure*}
\includegraphics[width=1\textwidth]{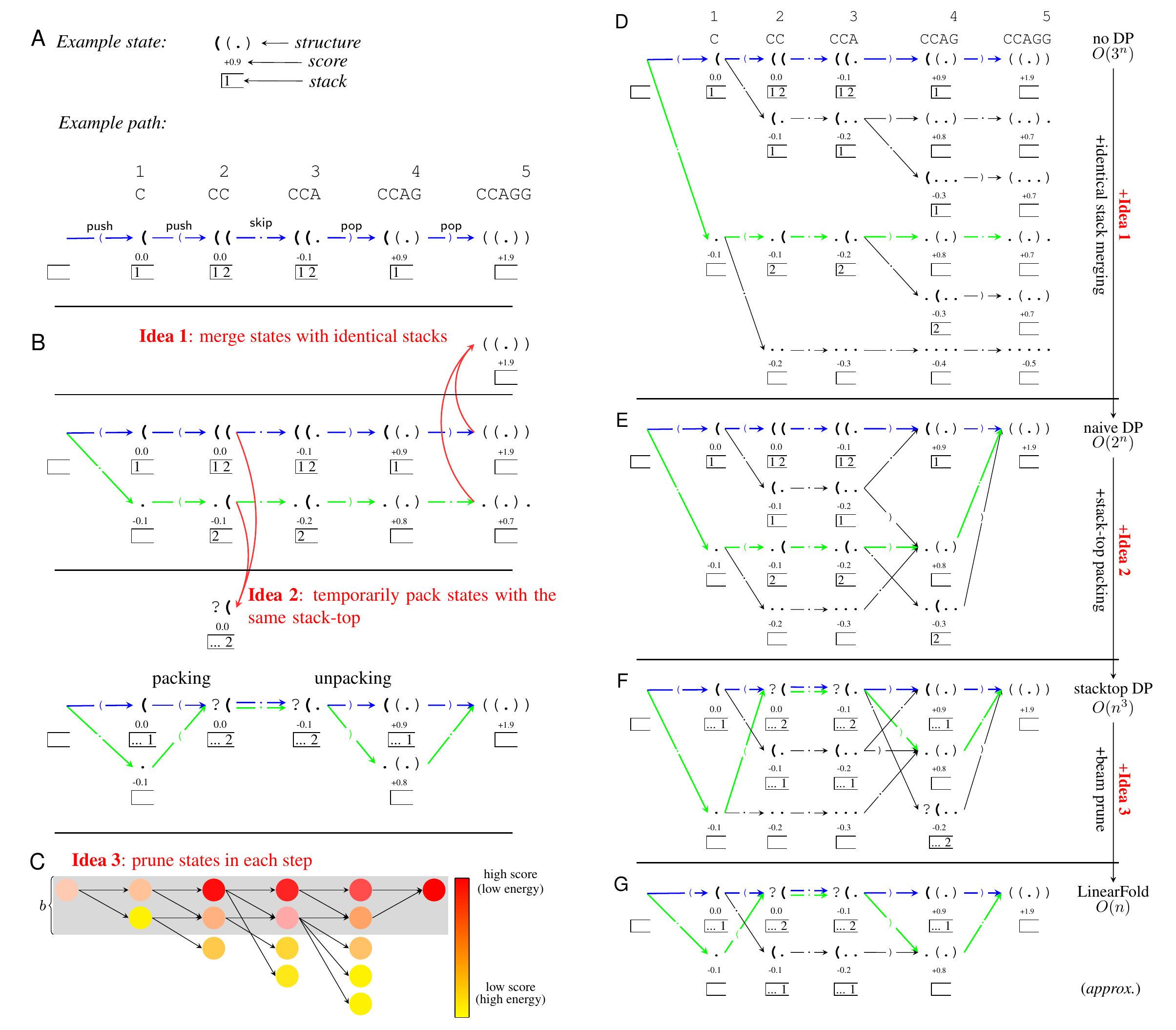}
\vspace{-0.8cm}
\caption{Illustration of the \linearfold approach,
  using a short sequence {\tt CCAGG} and the simple scoring function (Eq.~\ref{eq:simple}). 
  {\bf A}: an example state and an example (actually optimal) path, showing states (predicted prefix structures), actions  (\push ``\ml'', \nskip ``\md'', and \pop ``\mr''), and stacks (unpaired open brackets which are shown in bold in states).
  {\bf B}: two example paths (the optimal one in blue and a suboptimal one in green) and two essential ideas of left-to-right dynamic programming: 
  merging equivalent states with identical stacks (Idea 1)
  and packing temporarily equivalent states sharing the same stack top, and corresponding unpacking upon \pop (Idea 2). 
  {\bf C}: illustration of beam search, which keeps top $b$ states (those in the shaded region) per step (Idea 3).
  {\bf D}: the whole search space of the naive 
  algorithm ($O(3^n)$ time).
  {\bf E}: improving to $O(2^n)$ time with Idea 1.
  {\bf F}: further improving to $O(n^3)$ time with Idea 2.
  {\bf G}: further improving to $O(n)$ time (but with approximate search) with Idea 3. 
  In {\bf B, F}, and {\bf G}, each green/blue arrow pair \mq\bml {\raisebox{0.05cm}{\color{green}{\tiny$-\cdot\!\!\rightarrow$}}}\hspace{-.5cm}{\raisebox{-0.03cm}{\color{blue}{\tiny$-\cdot\!\!\rightarrow$}}} \mq\bml\md\/  is actually a single arrow, denoting two paths temporarily packed as one;
  we draw paired arrows 
  to highlight that two states \md\bml\/ and \bml\bml\/ are performing \nskip action together. 
  Note the version up to Idea 2 is exact and worst-case $O(n^3)$ time.
    \label{fig:method} }
    \vspace{-0.3cm}
\end{figure*}
\fi

\vspace{0.2cm}

\subsection{Idea 0: Brute-Force Search: $O(3^n)$}
\vspace{0.1cm}

The initial idea, introduced in 
Fig.~\ref{fig:overall}B, 
is to scan the RNA sequence left-to-right,
maintaining a {\em stack} along the way, and
performing one of the three actions
(push, skip, or pop)
at each step. 
More formally, we denote each {\em state} at step $j$ ($j=0...n$)
as a tuple along with a score $s$:
\[
        \nnitem{\vecy}{\sigma}{j}{s},
\]
where \vecy is the (sub)structure for the prefix $x_1 \ldots x_j$,
and
$\sigma$ is the  stack consisting of unmatched opening bracket positions in \vecy.
For example, in step 4, if \vecy=``\bml\ml\md\mr'', then $\sigma=[1]$ and $s=0.9$ (see Fig.~\ref{fig:method}A);
note that 
we denote open brackets in bold.
Each state at step $j$ can transition into a subsequent state of step $j\!+\!1$, taking one of the three
actions:
\begin{enumerate}
\item \push:
  label $x_{j+1}$ as ``\ml'' for it to be paired with a downstream nucleotide, and pushing $j+1$ on to the stack, 
  notated:
  \[
  \frac{\nnitem{\vecy}{\sigma}{j}{s}}{\nnitem{\vecy\!\circ\!{\text `}\ml{\text '}}{\sigma | (j+1)}{j+1}{s}}
  \]
\item \nskip:
  label $x_{j+1}$ as ``\md'' (unpaired and skipped): 
  \[
  \frac{\nnitem{\vecy}{\sigma}{j}{s}}{\nnitem{\vecy\!\circ\!{\text `}\md{\text '}}{\sigma}{j+1}{s + w_\unpaired}}
  \]
\item \pop:
  label $x_{j+1}$ as ``\mr'', paired with the upstream nucleotide $x_i$
  where $i$ is the top of the stack, and pop $i$
  (if $x_i x_{j+1}$ pair is allowed):
  \[
  \frac{\nnitem{\vecy}{\sigma | i}{j}{s}}{\nnitem{\vecy\!\circ\!{\text `}\mr{\text '}}{\sigma}{j+1}{s+ w_{x_i x_{j+1}}}}
  \]
\end{enumerate}

We start with the init state $\nnitem{\text{`'}}{[\,]}{0}{0}$ and finish with any state $\nnitem{\vecy}{[\,]}{n}{s}$ with an empty stack
(ensuring the output is a well-balanced dot-bracket sequence). 
See Fig.~\ref{fig:method}A for an example path for input sequence CCAGG, and Fig.~\ref{fig:method}D for all valid paths.

The above procedure describes a naive exhaustive search without dynamic programming
which has exponential runtime
$O(3^n)$, as there are up to three actions per step (see Fig.~\ref{fig:method}D).


Next, Fig.~\ref{fig:method}B
sketches
the two key dynamic programming ideas that speed up
this algorithm to $O(n^3)$ by merging and packing states.

\subsection{Idea 1 (DP): Merge States with Identical Stacks: $O(2^n)$}

We first observe that different states can have the same stack; for example,
in step 5, both ``\md\ml\md\mr\md'' and ``\ml\ml\md\mr\mr'' have the same empty stack (see Fig.~\ref{fig:method}B, Idea 1);
and in step 4, both ``\bml\md\md\md'' and ``\bml\ml\md\mr'' have the same stack [1] (see Fig.~\ref{fig:method}D).
These states can be merged, because even though they have different histories, going forward they are exactly equivalent.
After merging we save the state with the highest score and discard all others which have no potential to lead to the optimal structure.
More formally, we merge two states with the same stack:
\[
\begin{rcases}
  \nnitem{\vecy}{\sigma}{j}{s}\\
  \nnitem{\vecy'}{\sigma}{j}{s'}
\end{rcases}
\!\!\!\goesto \nitemt{\sigma}{j}{\vecy'', s''} 
\]
where 
\[
\tuple{\vecy'', s''} =
\begin{cases}
  \tuple{\vecy, s} & \text{if $s > s'$}\\
  \tuple{\vecy', s'} & \text{otherwise}
\end{cases}
\]

This algorithm is faster but still has exponential $O(2^n)$ time as there are exponentially many different stacks (see Fig.~\ref{fig:method}E).

\subsection{Idea 2 (DP): Pack Temporarily Equivalent States: $O(n^3)$}
\label{sec:idea2}

We further observe that even though some states have different stacks,
they might share the same stack top.
For example, 
in step 2, ``\md\bml'' and ``\bml\bml'' have [2] and [1,2] as their stacks, resp.,
but with the same stack top 2. 
Our key insight is that two states with the same stack-top
are ``temporarily equivalent'' and can be ``packed'' as they would behave equivalently until the
stack-top open bracket is closed (i.e., matched), after which they ``unpack'' and diverge.
As shown in Fig.~\ref{fig:method}B (Idea 2),
both ``\md\bml'' and ``\bml\bml'' are looking for a ``G'' to match with the stack top $x_2$=``C'',
and can be packed as ``\mq\bml'' with stack [...2] where \mq\/ and ``...'' represent
histories that are not important for now. After skipping the next nucleotide $x_3$=``A'', they become ``\mq\bml\md''
and upon matching the next nucleotide $x_4$=``G'' with the stack-top $x_2$=``C'', they unpack, resulting in ``\md\ml\md\mr'' and ``\bml\ml\md\mr''.

More formally, two states $\nitemt{\sigma | i}{i}{\vecy, s}$ and $\nitemt{\sigma' | i}{i}{\vecy', s'}$ sharing the same stack top can be packed:
\[
\begin{rcases}
  \nitemt{\sigma | i}{i}{\vecy,s}\\
  \nitemt{\sigma' | i}{i}{\vecy',s'}
\end{rcases}
\!\!\!\goesto \nitemt{i}{i}{\bml, 0} 
\]
Note that (a) we only need two indices to index the packed state;
(b) we omit the \mq's since they contain no information; and (c) somewhat counterintuitively, the resulting packed state's (sub)structure and score, $\tuple{\bml, 0}$
do not depend on the original states before packing.
More formally, for any packed state $\nitemt{i}{j}{\vecy, s}$,
its \vecy is a substructure only for the substring $x_i ... x_j$,
and its score $s$ is also for that portion only, i.e., $s=\scorew(x_i ... x_j, \vecy)$.
We can grow it by skip
\[
\frac{\nitemt{i}{j}{\vecy, s}}{\nitemt{i}{j+1}{\vecy\circ\text{`\md'}, s+\wunpaired}}
\]
or push actions
\[
\frac{\nitemt{i}{j}{\vecy, s}}{\nitemt{j+1}{j+1}{\bml, 0}}.
\]

The pop action is more involved. If $x_i$ and $x_{j+1}$ match,
we pop $i$, but where can we find the ``previous stack top''? It is {\em not} specified in the packed state.
Therefore, we need to find a state $\nitemt{k}{i-1}{\vecy', s'}$ that combines with the current state:
\[
\frac{\nitemt{k}{i-1}{\vecy',s'} \quad \nitemt{i}{j}{\vecy, s}}{\nitemt{k}{j+1}{\vecy'\circ\vecy\circ\text{`\mr'}, s'+s+w_{x_i  x_{j+1}}}}
\]

This version (see Fig.~\ref{fig:method}F) 
runs in worst-case $O(n^3)$ time, because the pop step involve three free indices.
It guarantees to return the optimal-scoring structure. 
It is inspired by a well-established algorithm in natural language parsing \citep{tomita:1988,huang+sagae:2010}; see Supplementary Fig.~\ref{fig:connect}.
Although this $O(n^3)$ runtime is the same as those classical bottom-up ones,
its unique left-to-right nature makes it amenable to $O(n)$ beam search. 

\subsection{Idea 3 (Approximate Search): Beam Pruning: $O(n)$}
\label{sec:idea3}

We further employ beam pruning~\cite{huang+:2012},
a popular heuristic widely used in computational linguistics,
to reduce the complexity from $O(n^3)$ to $O(n)$,
but with the cost of exact search.
Basically, at each step $j$, we only keep the $b$ top-scoring (lowest-energy) states 
and prune the other, less promising, ones (because they are less likely to be part of the optimal final structure).
This results in an approximate search algorithm in  $O(nb^2)$ time,
depicted in 
Figure~\ref{fig:method}C and G.
On top of beam search, we borrow $k$-best parsing
\cite{huang+chiang:2005}
to reduce the runtime to $O(n b\log b)$.
Here the beam size $b$ is a small constant (by default 100) so the overall runtime is linear in $n$.
We will show that 
our approximate search achieves even higher overall accuracy than the classical exact search methods.
The space complexity is $O(nb)$.
See Supplementary Fig.~\ref{fig:realdeduct} for the real system.
  There are two minor restrictions in our real system:
  the length of an interior loop is bounded by 30\nts
  (a standard limit found in most existing RNA folding software such as \contrafold),
  so is the leftmost (5'-end) unpaired segment of a multiloop
  (new constraint).
  These conditions are valid for 37\textdegree C, and no violations were found
  in the ArchiveII dataset.

\iftrue
\begin{figure}
  \centering
 \includegraphics[width=.49\textwidth]{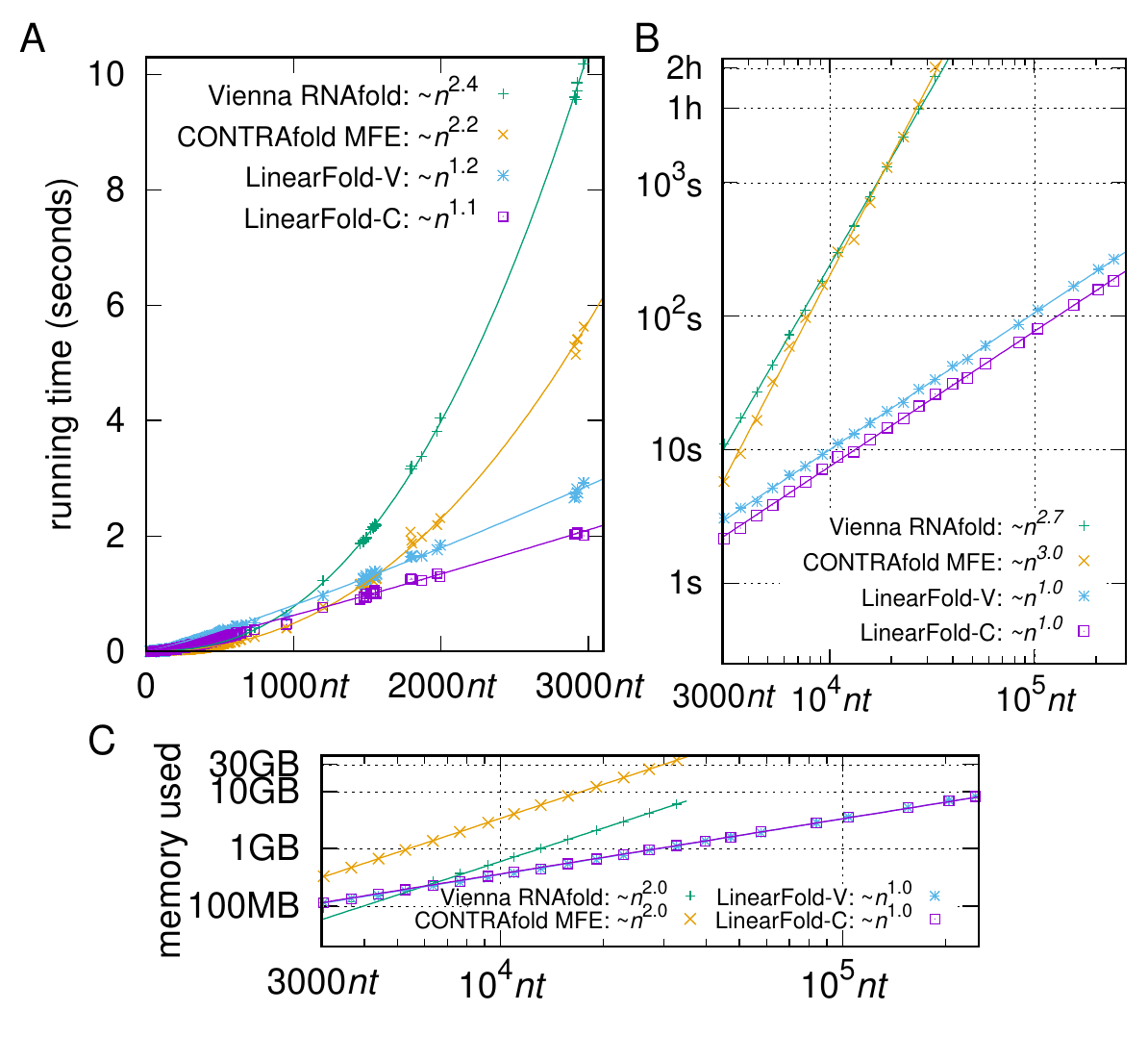}
 \vspace{-.7cm}
  \caption{Efficiency and scalability of \linearfold. 
    {\bf A}: runtime comparisons on the ArchiveII dataset
    with the two baselines, \contrafoldmfe and \viennarnafold. 
    {\bf B}: runtime comparisons on the RNAcentral dataset (log-log).
    {\bf C}: memory usage comparisons (RNAcentral set, log-log).
    \linearfold uses $O(n)$ time and memory, being substantially faster and slimmer than the $O(n^3)$-time, $O(n^2)$-space, baselines on long sequences.
    \label{fig:time}\vspace{-0.3cm}
  }
\end{figure} 
\fi

\iftrue
\begin{figure*}[!ht]
  \includegraphics[width=1\textwidth]{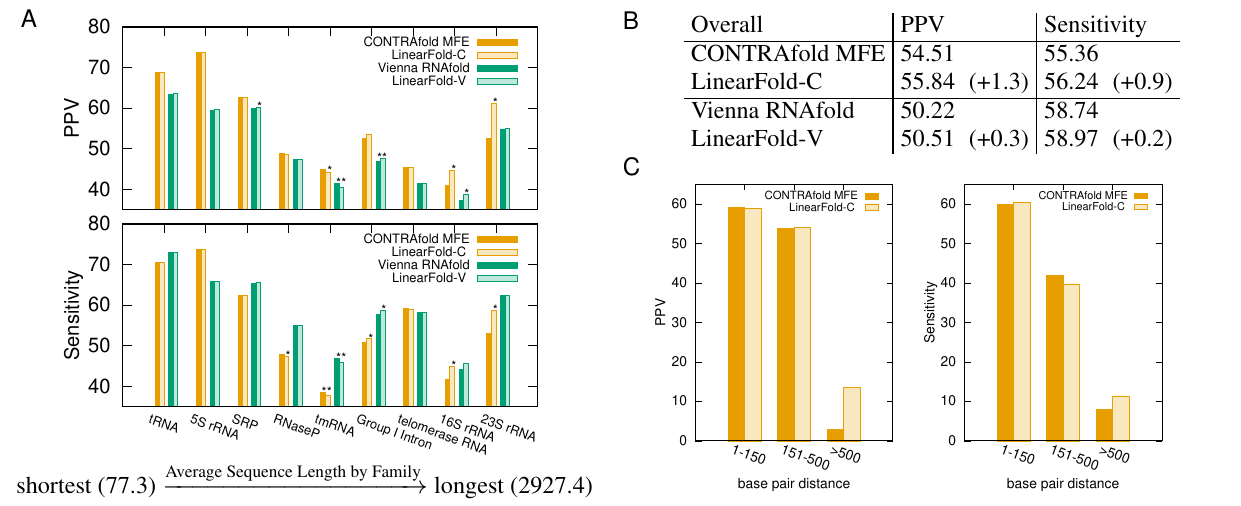} %
  \\[-0.6cm]
  \caption{Accuracy of \linearfold.
    {\bf A}: Each bar represents PPV/sensitivity
    averaged over all sequences in one family.
    Statistical significance is marked as 
    $^*$($0.01 \!\leq\! p \! < \! 0.05$) or $^{**}$($p\! <\! 0.01$). 
    See Table~\ref{tab:accuracy} for details. 
    {\bf B}: The overall accuracies, 
    averaging over all families.
   {\bf C}: 
   Each bar represents the overall PPV/sensitivity of all base pairs in a certain length range
   across all sequences.
 Supplementary Fig.~\ref{fig:vienna_distance} shows a similar result for \linearfoldv. 
   Overall, \linearfold outperforms exact search baselines, esp.~on longer families and long-range 
   pairs.
    \vspace{-0.4cm}    
    \label{fig:accuracy}}
\end{figure*}
\fi

\vspace{0.2cm}
\section{Results}

\vspace{0.1cm}

\subsection{Efficiency and Scalability} 

\vspace{0.1cm}

We compare \linearfold's efficiency with classical cubic-time algorithms
represented by 
\contrafold (Version 2.02) and \viennarnafold (Version 2.4.10) (\url{http://contra.stanford.edu/} and \url{https://www.tbi.univie.ac.at/RNA/download/sourcecode/2_4_x/ViennaRNA-2.4.10.tar.gz}).
We use two datasets: 
(a) the ArchiveII dataset \citep{sloma+mathews:2016},  a diverse
set of RNA sequences with known structures (\url{http://rna.urmc.rochester.edu/pub/archiveII.tar.gz};
we removed those sequences found in the S-Processed set. See Supplementary Table~\ref{tab:accuracy} for details),
and (b) a sampled subset of RNAcentral \citep{rnacentral:2017} (\url{https://rnacentral.org/}),
a comprehensive set of 
ncRNA
sequences from many databases.
While ArchiveII contains sequences of 3,000\nts or less,
RNAcentral has many much longer ones, with the longest being 244,296\nts 
(Homo Sapiens Transcript NONHSAT168677.1, from the NONCODE database \citep{noncode:2016}).
We run all programs (compiled by GCC 4.9.0) on Linux,
with 3.40GHz Intel Xeon E3-1231 CPU and 32G memory. 

Figure~\ref{fig:time}A shows that on the relatively short ArchiveII set, \linearfold's runtime
scales almost linearly with the sequence length,
while the two baselines  
have superquadratic runtimes.
On the much longer RNAcentral set,
Figure~\ref{fig:time}B 
shows strictly linear runtime for \linearfold and near-cubic runtimes for the baselines,
which 
agrees with the asymptotic analyses
and suggests that the minor deviations from the theoretical runtimes  
are due to the short sequence lengths in the ArchiveII set.
For a sequence of $\sim$10,000\nts
(e.g., the HIV genome),
\linearfold 
takes only 8 seconds
while the baselines take 4 minutes.
For a sequence of 32,753\nts,
\linearfold takes  26 seconds
while \contrafold and \rnafold take 2 and 1.7 hours, resp. 

\iftrue
\begin{figure*}[t]
  \centering
  \vspace{0.1cm}
    \includegraphics[width=1\textwidth]{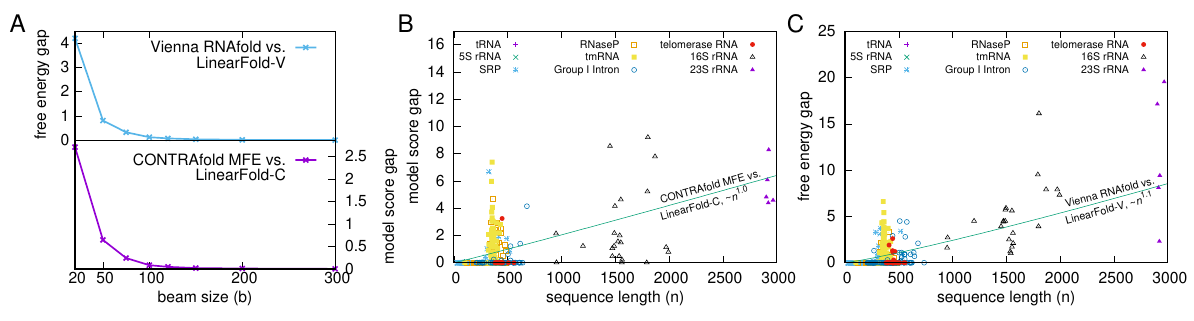} %
  \\[-0.3cm]
\caption{Search error (model score gap or free energy gap $\Delta\Delta G$). 
{\bf A}: average free energy gap (Vienna \rnafold vs.~\linearfoldv) and model cost gap (\contrafold vs.~\linearfoldc)
    with varying beam size;
    the search error 
    shrinks with beam size, quickly converging to 0.
    {\bf B} and {\bf C}: the search error (or gap) grows linearly with sequence length.
    Here tmRNA is the outlier with disproportionally severe search errors, which can explain
    the slightly worse accuracies of \linearfold on tmRNA in Fig.~\ref{fig:accuracy}A.
    See Supplementary Fig.~\ref{fig:searcherror-closeup} for a close-up on short sequences.
    \label{fig:searcherror}
}
\end{figure*}
\fi

In addition, \linearfold uses only $O(n)$ memory (Fig.~\ref{fig:time}C). 
The classical $O(n^3)$-time algorithm uses $O(n^2)$ space,
because 
it needs to solve the best-scoring 
substructure
for each substring $[i, j]$ bottom-up. 
\linearfold, by contrast, uses $O(n)$ space thanks to left-to-right beam search,
and is the first $O(n)$-space algorithm to be able to predict base pairs of unbounded distance.
It is able to fold the longest sequence in RNAcentral (244,296\nts)
within 3 minutes
while
neither \contrafold or \rnafold 
runs on anything longer than 32,767\nts
due to datastructure limitations.
As a result, the sequence limit on our web server ($10^5$\nts, see abstract) is
10x that of RNAfold web server (the previous largest),
being by far the largest limit
among all available servers (as of March 2019).
The curve-fittings in Fig.~\ref{fig:time}
were done log-log in {\small\tt gnuplot} 
with $n\!>\! 10^3$ in A,  $n \! >\! 3\!\times\! 10^3$ in B, and $n \! >\! 10^4$ in C,
to focus on the asymptotics.

\iftrue
\begin{figure*}
  \centering
    \includegraphics[width=1\textwidth]{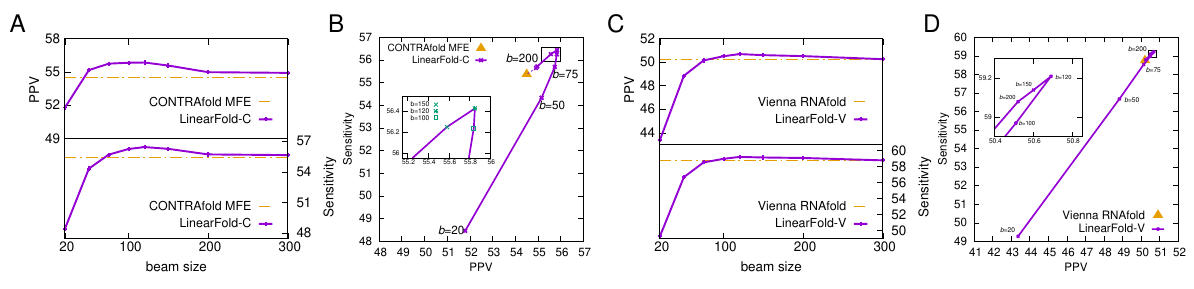} 
  \\[-0.3cm]
\caption{Impacts of beam size on prediction accuracy.
  {\bf A} and {\bf C}: PPV and Senstivity with varying beam size for \linearfoldc ({\bf A}) and
  \linearfoldv ({\bf C});
    {\bf B} and {\bf D}: PPV-sensitivity tradeoff for \linearfoldc ({\bf B}) and \linearfoldv ({\bf D}).
    Note that  \linearfold with $b=\infty$ is exact search in $O(n^3)$ time (Idea 2) and
    produces identical results to the baselines.
    \label{fig:beamsize}\vspace{-0.3cm}}
\end{figure*}
\fi

\subsection{Accuracy} 



We next compare
\linearfold
with the two baselines in accuracy,
reporting both Positive Predictive Value (PPV, 
the fraction of predicted pairs in the known structure) and sensitivity 
(the fraction of known pairs predicted)
on each RNA family in the ArchiveII dataset,
allowing correctly predicted pairs to be offset by one position for one nucleotide as compared to the known structure~\cite{sloma+mathews:2016};
we also report exact match accuracies in Supplementary Table~\ref{tab:accuracy_nos}.
We test statistical significance using a paired, one-sided
permutation test, following 
\citep{Aghaeepour+:2013}. 

%
%
%
%
%



Figure~\ref{fig:accuracy} shows that
\linearfold is more accurate than the baselines,
and interestingly, this advantage is more pronunced on longer sequences.
Individually, \linearfoldc (the \linearfold implementation of the \contrafold
model) is significantly more accurate in sensitivity than \contrafold on one family (Group I Intron), and both PPV/sensitivity
on two families (16S and 23S ribosomal RNAs), 
with the last two being the longest families in this dataset,
where they have average lengths 1548\nts and 2927\nts,
and enjoyed +3.56\%/+3.09\% and +8.65\%/+5.66\% (absolute) improvements in PPV/sensitivity, respectively.
\linearfoldv (the \linearfold implementation of the \viennarnafold
model) also outperforms \rnafold 
with significant improvements in PPV on two families (SRP and 16S rRNA), and both PPV/sensitivity on one family (Group I Intron).
Overall (across all families),
\linearfoldc outperforms \contrafold by
+1.3\%/+0.9\% PPV/sensitivity, 
while \linearfoldv outperforms \rnafold
by +0.3\%/+0.2\%.
See Supplementary Table~\ref{tab:accuracy} for details. 


Long-range base pairs are notoriously 
difficult to predict under current models~\cite{amman+:2013}.
Interestingly,
\linearfold is more accurate in both PPV and sensitivity than the exact search algorithm
for long-range base pairs of nucleotides greater than 500 nucleotides apart, as shown in
Fig.~\ref{fig:accuracy}C.
Combined with Supplementary Fig.~\ref{fig:vienna_distance},
we conclude that \linearfold
is more selective in predicting long-range base pairs (higher PPV),
but nevertheless predicts {\em more} such pairs that are correct
(higher Sensitivity).
Supplementary Fig.~\ref{fig:bylencnt}B--C further shows that both \linearfoldc and \linearfoldv
correct the severe overprediction of those long-range base pairs in exact search baselines.

\iftrue
\begin{figure}[!htb]
\begin{minipage}{\textwidth}
   \includegraphics[width=1\textwidth]{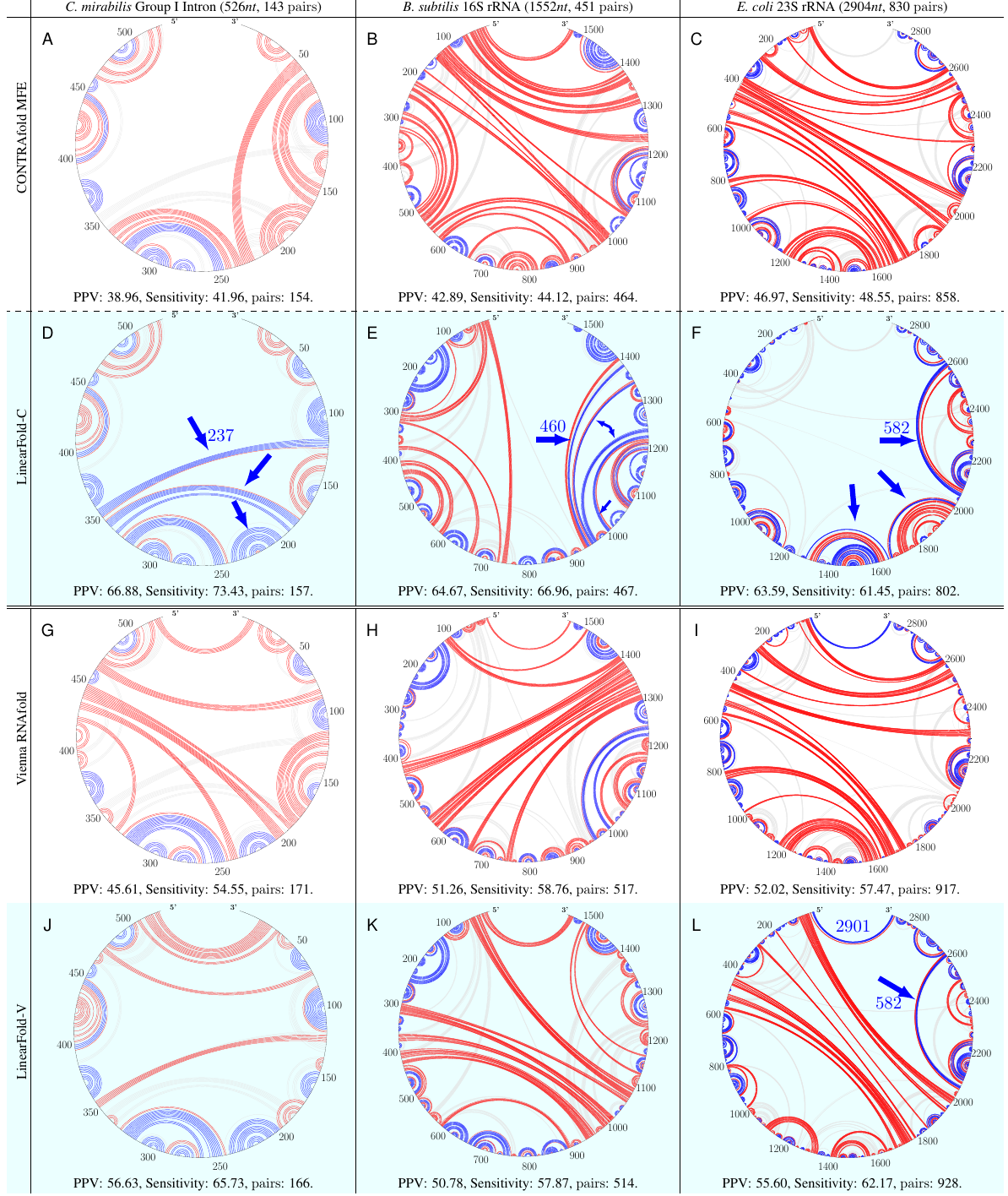} 
  \vspace{-.5cm}
  \captionof{figure}{Circular plots of the prediction results on three RNA sequences (from three different RNA families) comparing the baselines ({\bf A}--{\bf C}: \contrafoldmfe; {\bf G}--{\bf I}: \viennarnafold) and our \linearfold ({\bf D}--{\bf F}: \linearfoldc; {\bf J}--{\bf L}: \linearfoldv).
  Correctly predicted base pairs are in blue (true positives),
  incorrectly predicted pairs in red (false positives),
  and missing true base pairs in light gray (false negatives).
  Each plot is clockwise from 5' to 3'. 
  We can observe that
     (1) our \linearfold greatly reduces the false positives, esp.~on \contrafold;
     (2) our \linearfold correctly predicts many long-range pairs, e.g.,
     \linearfoldc on all three sequences ({\bf D}--{\bf F}) and \linearfoldv on  \ecoli 23S rRNA ({\bf L});
     (3) our \linearfold is able to predict the longest 5'-3' pairs, even with
     the beam size of 100, which is an order of magnitude smaller than the
     sequence lengths of 16S and 23S rRNAs. 
     (4) in almost all cases (except for \linearfoldv on \bsubtilis  16S rRNA ({\bf K})), \linearfold substantially outperforms the corresponding baseline.
  \label{fig:circular}}
  \end{minipage}
\end{figure}
\fi

Interestingly, even though our algorithm scans 5'-to-3', the
accuracy does not degrade toward the 3'-end, shown in Supplementary Fig.~\ref{fig:toward3'}.

\vspace{0.2cm}

\subsection{Search Quality}

\vspace{0.2cm}

Above we used  beam size 100. Now we
investigate the impacts of varying beam size. We first study its impact on
search quality. Since our search is approximate,
we quantify the notion of {\em search error} \cite{huang+sagae:2010} as the difference in score or free energy between \vecyhat, the optimal structure returned by exact search, 
and \vecybar, the 
one found by our linear-time beam search,
i.e.,
\[
\scorew(\vecx,\vecyhat) - \scorew(\vecx,\vecybar).
\]
The smaller this gap, 
the better the search quality.
Figure~\ref{fig:searcherror}A shows that
search error shrinks with beam size, 
quickly converging to 0 (exact search);
Figure~\ref{fig:searcherror}B--C show that
the search error (at $b=100$) grows linearly with sequence length,
indicating that our search quality does not degrade with longer sequences
(the average search error per nucleotide stays the same).

\vspace{0.2cm}

\subsection{Impacts of Beam Size on Prediction Accuracy}

\vspace{0.2cm}

Figure~\ref{fig:beamsize}A plots PPV and sensitivity as a function of beam size. 
\linearfoldc outperforms  \contrafoldmfe in both PPV and sensitivity 
with $b\geq$ 75 
and is stable 
with $b\in [100,150]$.
Figure~\ref{fig:beamsize}B shows the tradeoff between PPV and sensitivity.
Both PPV and sensitivity increase initially with beam size, culminating at $b$=120, and then decrease,  converging to exact search.
We do not tune the beam size on any dataset and use the round number of 100 as default.
Figures~\ref{fig:beamsize}C--D show a similar trend for \linearfoldv.

\vspace{0.2cm}
\subsection{Example Predictions: Group I Intron, 16S \& 23S rRNAs} 
\vspace{0.2cm}

Fig.~\ref{fig:circular} visualizes
the predicted secondary structures from three RNA
families:  {\em Cryptothallus~mirabilis}\xspace Group I Intron,  {\em Bacillus~subtilis}\xspace 16S rRNA, and {\em Escherichia~coli}\xspace  23S rRNA.
%
%
We observe that \linearfold substantially reduces false positives
(shown in red),
especially on the \contrafold model.
It also correctly predicts many (clusters of) long-range base pairs
(true positives, shown in blue),
e.g., in
\cmirabilis Group I Intron with \linearfoldc (Fig.~\ref{fig:circular}D, pair distance 237\nts),
\bsubtilis 16S rRNA with \linearfoldc (Fig.~\ref{fig:circular}E, pair distance 460\nts),
\ecoli 23S rRNA with both \linearfoldc and \linearfoldv (Figs.~\ref{fig:circular}F and ~\ref{fig:circular}L, pair distance 582\nts).
This reconfirms \linearfold's advantage in predicting
long-range base pairs shown in Fig.~\ref{fig:accuracy}C. 
Moreover, 
\linearfold is able to predict the longest 5'-3' pairs, as shown in
\ecoli 23S rRNA with \linearfoldv (Fig.~\ref{fig:circular}L, pair distance 2,901\nts).
In most cases (except \linearfoldv on \bsubtilis 16S rRNA, Fig.~\ref{fig:circular}K),
\linearfold
improves substantially over the corresponding baselines.
By contrast, local folding methods do not predict any long-range pairs,
shown in Fig.~\ref{fig:si-circle}.
We use \verb|rnafold --maxBPspan 150| for local folding,
and this limit of 150 is the largest default limit
in the local folding literature and softwares.


\iftrue
\begin{figure}[H]
\vspace{0.2cm}
  \includegraphics[width=.5\textwidth]{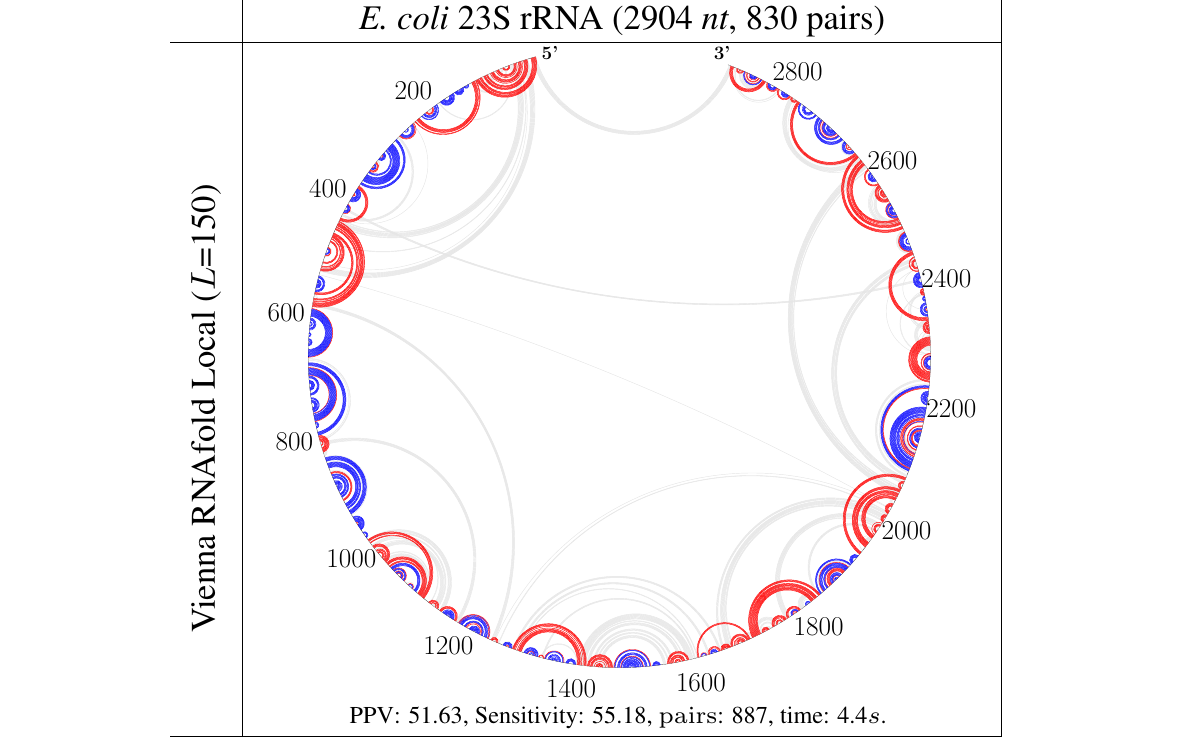}
 \vspace{-.2cm}
  \captionof{figure}{Circular plots of prediction results using the local folding mode of \viennarnafold
    (which only predicts local pairs no more than 150 \nts apart)
    on the \ecoli 23S rRNA  (corresponding to
    Figure~\ref{fig:circular}I). 
    Moreover, the  $O(nL^2)$-time local folding (with default $L=150$)
    is twice as slow as
    the $O(nb\log b)$-time \linearfoldv (with default $b=100$).
    \label{fig:si-circle}}
\end{figure} 
\fi

\section{Discussion}

There are several reasons why our beam search algorithm,
though approximate,
outperforms the exact search baselines in terms of accuracy (esp.~in 16S and 23S rRNAs and long-range base pairs).


\begin{enumerate}
\item  
First,
the scoring functions are imperfect, 
so it is totally possible for a suboptimal structure (in terms of model score or free energy)
to be more accurate than the optimal-score structure.
For example,
it was well studied that while the lowest free energy structure
contains only 72.9\% of the actual base pairs (given a dataset), 
a structure containing 86.1\% of them can be found with a free energy within 4.8\% of
the optimal structure \cite{zuker+:1991,mathews+:1999}.
\item
  Secondly, the beam search algorithm prunes lower-scoring (sub)structures at each step,
requiring the surviving (sub)structures and the final result to be highly scored for each prefix.
Our results suggest that this extra constraint,
like ``regularization'',
could compensate for
the inaccuracy of the (physical or machine-learning) model,
as \linearfold\/ {\em systematically} picks a more accurate
suboptimal structure without knowing the ground truth;
indeed, this seemingly surprising phenomenon has been observed before in computational linguistics  \citep{huang+sagae:2010}
which inspired this work.

\item
  Finally,
  our \linearfold algorithm resembles cotranscriptional folding
  where RNA molecules start to fold immediately before being fully transcribed
 \citep{gultyaev+:1995, meyer+miklos:2004}.
  This is  analogous to psycholinguistic evidence 
  that humans incrementally parse a sentence
  before it is fully read or heard \cite{frazier+rayner:1982}.
We hypothesize that some RNA sequences have evolved to fold
co-transcriptionally \cite{meyer+miklos:2004},
thus making our 5'-to-3' incremental 
approach 
more accurate than bottom-up baselines.
Supplementary Fig.~\ref{fig:3'-to-5'}B shows a slight preference for 5'-to-3' order over 3'-to-5'.
\end{enumerate}

There are other algorithmic efforts to speed up RNA folding,
including an $O(n^3/\log n)$ algorithm using the Four-Russians method \cite{venkatachalam+:2014},
and two other sub-cubic algorithms inspired by fast matrix multiplication and context-free parsing
\cite{zakov+:2011,bringmann+:2016}.
We note that all of them are based on the classical cubic-time bottom-up
algorithm,
and thus orthogonal to our left-to-right approach.
There also exists a linear-time algorithm~\cite{rastegari+condon:2005}
  to analyze a {\em given} structure, but not to predict one {\em de novo}.

\vspace{-0.3cm}

\section{Conclusion and Future Work}


We designed an $O(n)$-time, $O(n)$-space, approximate search algorithm,
using incremental dynamic programming plus beam search,
and apply this algorithm to both machine-learned and thermodynamic models.
Besides the linearity in both time and memory (Fig.~\ref{fig:time}),
we also found:
\begin{enumerate}
\item Though \linearfold uses only a fraction of time and memory compared to existing algorithms,
  our predicted structures are  even more accurate overall in both PPV and sensitivity
  and on both machine-learned and thermodynamic models
      (see Fig.~\ref{fig:accuracy}).
\item The accuracy improvement of \linearfold is more pronunced on longer families such as 16S and 23S rRNAs (see Figs.~\ref{fig:accuracy} and~\ref{fig:circular}).
\item 
  \linearfold is also more accurate than the baselines 
      at predicting long-range base pairs over 500\nts apart
      (Figs.~\ref{fig:accuracy}C), 
      which is well known to be challenging for the current models \cite{amman+:2013}.
\item Although the performance of \linearfold depends on the beam size $b$, the number of base pairs and
  the accuracy of prediction are stable when $b$ is in the range of 100--200. 
\end{enumerate}

There is a crucial difference between our \linearfold
and local folding algorithms \citep{kiryu+:2008,bernhart+:2006, Lange+:2012}
that can only predict pairs up to a certain distance.
Theoretical and empirical studies found 
several evidences that {\em unboundedly} long-distance pairs 
are actually quite common in natural RNA structures:
(a) the length of the longest 
base pair 
grows nearly linearly with sequence length $n$ \cite{li+reidys:2018};
(b) the physical distance between the 5'--3' ends in folded structures
is short and nearly constant \cite{lai+:2018,yoffe+:2011,leija+:2014}.


Our work has several potential extensions.
\begin{enumerate}
\item It is possible that \linearfold can be extended to calculate the partition
  function and base pair probabilities for natural RNA sequences with well-defined structures,
since the classical method for that task, the McCaskill (\citeyear{mccaskill:1990}) algorithm, 
is isomorphic in structure to the cubic-time 
algorithms that are used as baselines in this paper.
%
%
\item This linear-time approach to calculate base pair probabilities should facilitate the linear-time identification of pseudoknots,
by either replacing the cubic-time McCaskill algorithm with a linear-time one in those
heuristic pseudoknot-prediction programs \cite{bellaousov+mathews:2010,sato+:2011},
or linearizing a supercubic-time dynamic program for direct
 prediction with pseudoknots \citep{dirks+pierce:2003, reeder+giegerich:2004}. 

\item We will test the hypothesis that our beams  potentially
  capture  cotranscriptional folding  with empirical data on cotranscriptional folding \cite{watters+:2016}.
  
\item Being linear-time, \linearfold also facilitates faster parameter training than the cubic-time \contrafold
using structured prediction methods \cite{huang+:2012}, and 
we envision a more accurate \linearfold using a model tailored to its own search. 
\end{enumerate}

\vspace{-0.3cm}

\section*{Acknowledgements}

We would like to thank the reviewers for suggestions, Rhiju Das for encouragement and early adoption of LinearFold into the EteRNA game, James Cross for help in algorithm design, 
and Juneki Hong and Liang Zhang for proofreading.

\vspace{-0.3cm}

\section*{Author Contributions}

L.H.~conceived the idea and directed the project. 
L.H., D.D., and K.Z.~designed algorithms.
L.H.~and D.D.~wrote a Python prototype,
and K.Z., D.D., and H.Z.~wrote the fast C++ version.
D.H.M.~and D.H.~guided the evaluation 
that H.Z.~and D.D.~carried out. 
L.H., D.D., and H.Z.~wrote the manuscript; D.H.M.~and D.H.~revised it.
K.L.~made the webserver. 

\vspace{-0.3cm}
\section*{Funding}

This project was supported in part by
National Science Foundation [IIS-1656051 and IIS-1817231 to L.H.],
National Institutes of Health [R56 AG053460 and R21 AG052950 to D.H., and R01 GM076485 to D.H.M.].

\vspace{-0.3cm}

\balance
\bibliographystyle{natbib}
\bibliography{LinearFold}

\begin{thebibliography}{}

\bibitem[Aghaeepour and Hoos(2013)Aghaeepour and Hoos]{Aghaeepour+:2013}
Aghaeepour, N. and Hoos, H.~H. (2013).
\newblock Ensemble-based prediction of {RNA} secondary structures.
\newblock {\em BMC bioinformatics\/}, {\bf 14}(139), 1.

\bibitem[Amman {\em et~al.}(2013)Amman, Bernhart, Doose, Hofacker, Qin,
  Stadler, and Will]{amman+:2013}
Amman, F., Bernhart, S.~H., Doose, G., Hofacker, I.~L., Qin, J., Stadler,
  P.~F., and Will, S. (2013).
\newblock The trouble with long-range base pairs in {RNA} folding.
\newblock In J.~Setubal and N.~Almeida, editors, {\em Proceedings of the 8th
  Brazilian Symposium on Bioinformatics\/}, pages 1--11, Recife, Brazil.
  Springer International Publishing Switzerland.

\bibitem[Andronescu {\em et~al.}(2007)Andronescu, Condon, Hoos, Mathews, and
  Murphy]{andronescu+:2007}
Andronescu, M., Condon, A., Hoos, H.~H., Mathews, D.~H., and Murphy, K.~P.
  (2007).
\newblock Efficient parameter estimation for {RNA} secondary structure
  prediction.
\newblock {\em Bioinformatics\/}, {\bf 23}(13), i19--i28.

\bibitem[Angelbello {\em et~al.}(2018)Angelbello, Chen, Childs-Disney, Zhang,
  Wang, and Disney]{angelbello+:2018}
Angelbello, A.~J., Chen, J.~L., Childs-Disney, J.~L., Zhang, P., Wang, Z.-F.,
  and Disney, M.~D. (2018).
\newblock Using genome sequence to enable the design of medicines and chemical
  probes.
\newblock {\em Chemical reviews\/}, {\bf 118}(4), 1599--1663.

\bibitem[Bellaousov and Mathews(2010)Bellaousov and
  Mathews]{bellaousov+mathews:2010}
Bellaousov, S. and Mathews, D.~H. (2010).
\newblock Probknot: fast prediction of {RNA} secondary structure including
  pseudoknots.
\newblock {\em RNA\/}, {\bf 16}(10), 1870--1880.

\bibitem[Bernhart {\em et~al.}(2006)Bernhart, Hofacker, and
  Stadler]{bernhart+:2006}
Bernhart, S.~H., Hofacker, I.~L., and Stadler, P.~F. (2006).
\newblock Local {RNA} base pairing probabilities in large sequences.
\newblock {\em Bioinformatics\/}, {\bf 22}(5), 614--615.

\bibitem[Bringmann {\em et~al.}(2016)Bringmann, Grandoni, Saha, and
  Williams]{bringmann+:2016}
Bringmann, K., Grandoni, F., Saha, B., and Williams, V.~V. (2016).
\newblock Truly sub-cubic algorithms for language edit distance and
  {RNA}-folding via fast bounded-difference min-plus product.
\newblock In {\em 2016 IEEE 57th Annual Symposium on Foundations of Computer
  Science (FOCS)\/}, pages 375--384, New Brunswick, NJ, USA. IEEE.

\bibitem[Castanotto and Rossi(2009)Castanotto and Rossi]{castanotto+rossi:2009}
Castanotto, D. and Rossi, J.~J. (2009).
\newblock The promises and pitfalls of {RNA}-interference-based therapeutics.
\newblock {\em Nature\/}, {\bf 457}(7228), 426--433.

\bibitem[Childs-Disney {\em et~al.}(2007)Childs-Disney, Wu, Pushechnikov,
  Aminova, and Disney]{childs-disney+:2007}
Childs-Disney, J.~L., Wu, M., Pushechnikov, A., Aminova, O., and Disney, M.~D.
  (2007).
\newblock A small molecule microarray platform to select {RNA} internal
  loop-ligand interactions.
\newblock {\em ACS Chemical Biology\/}, {\bf 2}(11), 745--754.

\bibitem[Crooke(2004)Crooke]{crooke:2004}
Crooke, S. (2004).
\newblock Antisense strategies.
\newblock {\em Current Molecular Medicine\/}, {\bf 4}(5), 465--487.

\bibitem[Dirks and Pierce(2003)Dirks and Pierce]{dirks+pierce:2003}
Dirks, R.~M. and Pierce, N.~A. (2003).
\newblock A partition function algorithm for nucleic acid secondary structure
  including pseudoknots.
\newblock {\em Journal of computational chemistry\/}, {\bf 24}(13), 1664--1677.

\bibitem[Do {\em et~al.}(2006)Do, Woods, and Batzoglou]{do+:2006}
Do, C., Woods, D., and Batzoglou, S. (2006).
\newblock Contrafold: {RNA} secondary structure prediction without
  physics-based models.
\newblock {\em Bioinformatics\/}, {\bf 22}(14), e90--e98.

\bibitem[Eddy(2001)Eddy]{eddy:2001}
Eddy, S.~R. (2001).
\newblock Non-coding {RNA} genes and the modern {RNA} world.
\newblock {\em Nature Reviews Genetics\/}, {\bf 2}(12), 919--929.

\bibitem[Frazier and Rayner(1982)Frazier and Rayner]{frazier+rayner:1982}
Frazier, L. and Rayner, K. (1982).
\newblock Making and correcting errors during sentence comprehension: Eye
  movements in the analysis of structurally ambiguous sentences.
\newblock {\em Cognitive psychology\/}, {\bf 14}(2), 178--210.

\bibitem[Fu {\em et~al.}(2015)Fu, Xu, Lu, Zhao, and Mathews]{fu+:2015}
Fu, Y., Xu, Z.~Z., Lu, Z.~J., Zhao, S., and Mathews, D.~H. (2015).
\newblock Discovery of novel nc{RNA} sequences in multiple genome alignments on
  the basis of conserved and stable secondary structures.
\newblock {\em PloS One\/}, {\bf 10}(6), e0130200.

\bibitem[Gareiss {\em et~al.}(2008)Gareiss, Sobczak, McNaughton, Palde,
  Thornton, and Miller]{gareiss+:2008}
Gareiss, P.~C., Sobczak, K., McNaughton, B.~R., Palde, P.~B., Thornton, C.~A.,
  and Miller, B.~L. (2008).
\newblock Dynamic combinatorial selection of molecules capable of inhibiting
  the ({CUG}) repeat {RNA}-{MBNL1} interaction in vitro: discovery of lead
  compounds targeting myotonic dystrophy ({DM1}).
\newblock {\em Journal of the American Chemical Society\/}, {\bf 130}(48),
  16254--16261.

\bibitem[Gilbert(1986)Gilbert]{gilbert:1986}
Gilbert, W. (1986).
\newblock Origin of life: The {RNA} world.
\newblock {\em Nature\/}, {\bf 319}(6055).

\bibitem[Gruber {\em et~al.}(2010)Gruber, Findeiss, Washietl, Hofacker, and
  Stadler]{gruber+:2010}
Gruber, A., Findeiss, S., Washietl, S., Hofacker, I., and Stadler, P.~F.
  (2010).
\newblock {RNA}z 2.0: improved noncoding {RNA} detection.
\newblock In {\em Pacific Symposium on Biocomputing\/}, volume~15, pages
  69--79. World Scientific Publishing.

\bibitem[Gultyaev {\em et~al.}(1995)Gultyaev, Van~Batenburg, and
  Pleij]{gultyaev+:1995}
Gultyaev, A.~P., Van~Batenburg, F., and Pleij, C.~W. (1995).
\newblock The computer simulation of {RNA} folding pathways using a genetic
  algorithm.
\newblock {\em Journal of molecular biology\/}, {\bf 250}(1), 37--51.

\bibitem[Hofacker and Lorenz(2014)Hofacker and Lorenz]{hofacker+lorenz:2014}
Hofacker, I.~L. and Lorenz, R. (2014).
\newblock {\em Predicting {RNA} structure: advances and limitations\/}.
\newblock Humana Press, Totowa, NJ, USA.

\bibitem[Huang and Chiang(2005)Huang and Chiang]{huang+chiang:2005}
Huang, L. and Chiang, D. (2005).
\newblock Better $k$-best {P}arsing.
\newblock In {\em Proceedings of the Ninth International Workshop on Parsing
  Technologies (IWPT-2005)\/}, pages 53--64. {ACL}.

\bibitem[Huang and Sagae(2010)Huang and Sagae]{huang+sagae:2010}
Huang, L. and Sagae, K. (2010).
\newblock Dynamic programming for linear-time incremental parsing.
\newblock In {\em Proceedings of ACL 2010\/}, page 1077–1086, Uppsala,
  Sweden. {ACL}.

\bibitem[Huang {\em et~al.}(2012)Huang, Fayong, and Guo]{huang+:2012}
Huang, L., Fayong, S., and Guo, Y. (2012).
\newblock Structured perceptron with inexact search.
\newblock In {\em Proceedings of NAACL 2012\/}, pages 142--151. {ACL}.

\bibitem[Joyce(1994)Joyce]{joyce:1994}
Joyce, G.~F. (1994).
\newblock In vitro evolution of nucleic acids.
\newblock {\em Current opinion in structural biology\/}, {\bf 4}(3), 331--336.

\bibitem[Kasami(1965)Kasami]{kasami:1965}
Kasami, T. (1965).
\newblock An efficient recognition and syntax analysis algorithm for
  context-free languages.
\newblock Technical Report AFCRL-65-758, AFCRL.

\bibitem[Kiryu {\em et~al.}(2008)Kiryu, Kin, and Asai]{kiryu+:2008}
Kiryu, H., Kin, T., and Asai, K. (2008).
\newblock Rfold: an exact algorithm for computing local base pairing
  probabilities.
\newblock {\em Bioinformatics\/}, {\bf 24}(3), 367--373.

\bibitem[Knuth(1965)Knuth]{knuth:1965}
Knuth, D. (1965).
\newblock On the translation of languages from left to right.
\newblock {\em Information and Control\/}, {\bf 8}, 607--639.

\bibitem[Lai {\em et~al.}(2018)Lai, Kayedkhordeh, Cornell, Farah, Bellaousov,
  Rietmeijer, Mathews, and Ermolenko]{lai+:2018}
Lai, W.-J.~C., Kayedkhordeh, M., Cornell, E.~V., Farah, E., Bellaousov, S.,
  Rietmeijer, R., Mathews, D.~H., and Ermolenko, D.~N. (2018).
\newblock The formation of intramolecular secondary structure brings {mRNA}
  ends in close proximity.
\newblock {\em Nature Communications\/}, {\bf 9}(1), 4328.

\bibitem[Lange {\em et~al.}(2012)Lange, Maticzka, M{\"o}hl, Gagnon, Brown, and
  Backofen]{Lange+:2012}
Lange, S.~J., Maticzka, D., M{\"o}hl, M., Gagnon, J.~N., Brown, C.~M., and
  Backofen, R. (2012).
\newblock Global or local? predicting secondary structure and accessibility in
  {mRNAs}.
\newblock {\em Nucleic Acids Research\/}, {\bf 40}(12), 5215--5226.

\bibitem[Leija-Mart{\'\i}nez {\em et~al.}(2014)Leija-Mart{\'\i}nez,
  Casas-Flores, Cadena-Nava, Roca, Mendez-Caba{\~n}as, Gomez, and
  Ruiz-Garcia]{leija+:2014}
Leija-Mart{\'\i}nez, N., Casas-Flores, S., Cadena-Nava, R.~D., Roca, J.~A.,
  Mendez-Caba{\~n}as, J.~A., Gomez, E., and Ruiz-Garcia, J. (2014).
\newblock The separation between the 5'-3' ends in long {RNA} molecules is
  short and nearly constant.
\newblock {\em Nucleic Acids Research\/}, {\bf 42}(22), 13963--13968.

\bibitem[Li and Reidys(2018)Li and Reidys]{li+reidys:2018}
Li, T.~J. and Reidys, C.~M. (2018).
\newblock The rainbow spectrum of {RNA} secondary structures.
\newblock {\em Bulletin of mathematical biology\/}, {\bf 80}(6), 1514--1538.

\bibitem[Licon {\em et~al.}(2010)Licon, Taufer, Leung, and
  Johnson]{licon+:2010}
Licon, A., Taufer, M., Leung, M.-Y., and Johnson, K.~L. (2010).
\newblock A dynamic programming algorithm for finding the optimal segmentation
  of an {RNA} sequence in secondary structure predictions.
\newblock In {\em 2nd International Conference on Bioinformatics and
  Computational Biology\/}, pages 165--170. {ACM}.

\bibitem[Lorenz {\em et~al.}(2011)Lorenz, Bernhart, Zu~Siederdissen, Tafer,
  Flamm, Stadler, and Hofacker]{lorenz+:2011}
Lorenz, R., Bernhart, S.~H., Zu~Siederdissen, C.~H., Tafer, H., Flamm, C.,
  Stadler, P.~F., and Hofacker, I.~L. (2011).
\newblock Vienna{RNA} package 2.0.
\newblock {\em Algorithms for Molecular Biology\/}, {\bf 6}(1), 1.

\bibitem[Lu and Mathews(2008)Lu and Mathews]{lu+mathews:2008}
Lu, Z.~J. and Mathews, D.~H. (2008).
\newblock Efficient si{RNA} selection using hybridization thermodynamics.
\newblock {\em Nucleic Acids Research\/}, {\bf 36}(2), 640--647.

\bibitem[Mathews and Turner(2006)Mathews and Turner]{mathews+turner:2006}
Mathews, D.~H. and Turner, D.~H. (2006).
\newblock Prediction of {RNA} secondary structure by free energy minimization.
\newblock {\em Current Opinion in Structural Biology\/}, {\bf 16}(3), 270--278.

\bibitem[Mathews {\em et~al.}(1999)Mathews, Sabina, Zuker, and
  Turner]{mathews+:1999}
Mathews, D.~H., Sabina, J., Zuker, M., and Turner, D.~H. (1999).
\newblock Expanded sequence dependence of thermodynamic parameters improves
  prediction of {RNA} secondary structure.
\newblock {\em Journal of molecular biology\/}, {\bf 288}(5), 911--940.

\bibitem[Mathews {\em et~al.}(2004)Mathews, Disney, Childs, Schroeder, Zuker,
  and Turner]{mathews+:2004}
Mathews, D.~H., Disney, M.~D., Childs, J.~L., Schroeder, S.~J., Zuker, M., and
  Turner, D.~H. (2004).
\newblock Incorporating chemical modification constraints into a dynamic
  programming algorithm for prediction of {RNA} secondary structure.
\newblock {\em Proceedings of the National Academy of Sciences of the United
  States of America\/}, {\bf 101}(19), 7287--7292.

\bibitem[McCaskill(1990)McCaskill]{mccaskill:1990}
McCaskill, J.~S. (1990).
\newblock The equilibrium partition function and base pair binding
  probabilities for {RNA} secondary structure.
\newblock {\em Biopolymers\/}, {\bf 29}(6-7), 1105--1119.

\bibitem[Meyer and Miklos(2004)Meyer and Miklos]{meyer+miklos:2004}
Meyer, I.~M. and Miklos, I. (2004).
\newblock Co-transcriptional folding is encoded within {RNA} genes.
\newblock {\em BMC molecular biology\/}, {\bf 5}(1), 10.

\bibitem[Nussinov {\em et~al.}(1978)Nussinov, Pieczenik, Griggs, and
  Kleitman]{nussinov+:1978}
Nussinov, R., Pieczenik, G., Griggs, J.~R., and Kleitman, D.~J. (1978).
\newblock Algorithms for loop matchings.
\newblock {\em SIAM Journal on Applied mathematics\/}, {\bf 35}(1), 68--82.

\bibitem[Palde {\em et~al.}(2010)Palde, Ofori, Gareiss, Lerea, and
  Miller]{palde+:2010}
Palde, P.~B., Ofori, L.~O., Gareiss, P.~C., Lerea, J., and Miller, B.~L.
  (2010).
\newblock Strategies for recognition of stem-loop {RNA} structures by synthetic
  ligands: Application to the {HIV-1} frameshift stimulatory sequence.
\newblock {\em Journal of Medicinal Chemistry\/}, {\bf 53}(16), 6018--6027.

\bibitem[Rastegari and Condon(2005)Rastegari and Condon]{rastegari+condon:2005}
Rastegari, B. and Condon, A. (2005).
\newblock Linear time algorithm for parsing {RNA} secondary structure.
\newblock In {\em International Workshop on Algorithms in Bioinformatics\/},
  pages 341--352. Springer.

\bibitem[Reeder and Giegerich(2004)Reeder and Giegerich]{reeder+giegerich:2004}
Reeder, J. and Giegerich, R. (2004).
\newblock Design, implementation and evaluation of a practical pseudoknot
  folding algorithm based on thermodynamics.
\newblock {\em BMC bioinformatics\/}, {\bf 5}(1), 1.

\bibitem[Sato {\em et~al.}(2009)Sato, Hamada, Asai, and Mituyama]{sato+:2009}
Sato, K., Hamada, M., Asai, K., and Mituyama, T. (2009).
\newblock Centroidfold: a web server for {RNA} secondary structure prediction.
\newblock {\em Nucleic Acids Research\/}, {\bf 37}(suppl\_2), W277--W280.

\bibitem[Sato {\em et~al.}(2011)Sato, Kato, Hamada, Akutsu, and
  Asai]{sato+:2011}
Sato, K., Kato, Y., Hamada, M., Akutsu, T., and Asai, K. (2011).
\newblock Ipknot: fast and accurate prediction of {RNA} secondary structures
  with pseudoknots using integer programming.
\newblock {\em Bioinformatics\/}, {\bf 27}(13), i85--i93.

\bibitem[Sazani {\em et~al.}(2002)Sazani, Gemignani, Kang, Maier, Manoharan,
  Persmark, Bortner, and Kole]{sazani+:2002}
Sazani, P., Gemignani, F., Kang, S.-H., Maier, M., Manoharan, M., Persmark, M.,
  Bortner, D., and Kole, R. (2002).
\newblock Systemically delivered antisense oligomers upregulate gene expression
  in mouse tissues.
\newblock {\em Nature biotechnology\/}, {\bf 20}(12), 1228--1233.

\bibitem[Seetin and Mathews(2012)Seetin and Mathews]{seetin+mathews:2012}
Seetin, M.~G. and Mathews, D.~H. (2012).
\newblock {\em {RNA} structure prediction: an overview of methods\/}.
\newblock Humana Press, Totowa, NJ, USA.

\bibitem[Sloma and Mathews(2016)Sloma and Mathews]{sloma+mathews:2016}
Sloma, M. and Mathews, D. (2016).
\newblock Exact calculation of loop formation probability identifies folding
  motifs in {RNA} secondary structures.
\newblock {\em {RNA}, 22, 1808--1818\/}.

\bibitem[Stephens {\em et~al.}(2015)Stephens, Lee, Faghri, Campbell, Zhai,
  Efron, Iyer, Schatz, Sinha, and Robinson]{stephens+:2015}
Stephens, Z.~D., Lee, S.~Y., Faghri, F., Campbell, R.~H., Zhai, C., Efron,
  M.~J., Iyer, R., Schatz, M.~C., Sinha, S., and Robinson, G.~E. (2015).
\newblock Big data: astronomical or genomical?
\newblock {\em PLoS Biology\/}, {\bf 13}(7), e1002195.

\bibitem[Tafer {\em et~al.}(2008)Tafer, Ameres, Obernosterer, Gebeshuber,
  Schroeder, Martinez, and Hofacker]{tafer+:2008}
Tafer, H., Ameres, S.~L., Obernosterer, G., Gebeshuber, C.~A., Schroeder, R.,
  Martinez, J., and Hofacker, I.~L. (2008).
\newblock The impact of target site accessibility on the design of effective
  {siRNAs}.
\newblock {\em Nature biotechnology\/}, {\bf 26}(5), 578--583.

\bibitem[{The RNAcentral Consortium}(2017){The RNAcentral
  Consortium}]{rnacentral:2017}
{The RNAcentral Consortium} (2017).
\newblock {RNA}central: a comprehensive database of non-coding {RNA} sequences.
\newblock {\em Nucleic Acids Research\/}, {\bf 45}(D1), D128--D134.

\bibitem[Tomita(1988)Tomita]{tomita:1988}
Tomita, M. (1988).
\newblock Graph-structured stack and natural language parsing.
\newblock In {\em Proceedings of ACL\/}, page 249–257. {ACL}.

\bibitem[Venkatachalam {\em et~al.}(2014)Venkatachalam, Gusfield, and
  Frid]{venkatachalam+:2014}
Venkatachalam, B., Gusfield, D., and Frid, Y. (2014).
\newblock Faster algorithms for {RNA}-folding using four-russians method.
\newblock {\em Algorithms for Molecular Biology\/}, {\bf 9}(1), 5.

\bibitem[Washietl {\em et~al.}(2012)Washietl, Will, Hendrix, Goff, Rinn,
  Berger, and Kellis]{washietl+:2012}
Washietl, S., Will, S., Hendrix, D.~A., Goff, L.~A., Rinn, J.~L., Berger, B.,
  and Kellis, M. (2012).
\newblock Computational analysis of noncoding {RNA}s.
\newblock {\em Wiley Interdisciplinary Reviews: {RNA}\/}, {\bf 3}(6), 759--778.

\bibitem[Watters {\em et~al.}(2016)Watters, Strobel, Angela, Lis, and
  Lucks]{watters+:2016}
Watters, K.~E., Strobel, E.~J., Angela, M.~Y., Lis, J.~T., and Lucks, J.~B.
  (2016).
\newblock Cotranscriptional folding of a riboswitch at nucleotide resolution.
\newblock {\em Nature structural \& molecular biology\/}, {\bf 23}(12), 1124.

\bibitem[Watts {\em et~al.}(2009)Watts, Dang, Gorelick, Leonard, Bess~Jr,
  Swanstrom, Burch, and Weeks]{watts+:2009}
Watts, J.~M., Dang, K.~K., Gorelick, R.~J., Leonard, C.~W., Bess~Jr, J.~W.,
  Swanstrom, R., Burch, C.~L., and Weeks, K.~M. (2009).
\newblock Architecture and secondary structure of an entire {HIV-1} {RNA}
  genome.
\newblock {\em Nature\/}, {\bf 460}(7256), 711--716.

\bibitem[Yoffe {\em et~al.}(2011)Yoffe, Prinsen, Gelbart, and
  Ben-Shaul]{yoffe+:2011}
Yoffe, A.~M., Prinsen, P., Gelbart, W., and Ben-Shaul, A. (2011).
\newblock The ends of a large {RNA} molecule are necessarily close.
\newblock {\em Nucleic Acids Research\/}, {\bf 39}(1), 292--299.

\bibitem[Younger(1967)Younger]{younger:1967}
Younger, D.~H. (1967).
\newblock Recognition and parsing of context-free languages in time $n^3$.
\newblock {\em Information and Control\/}, {\bf 10}, 189--208.

\bibitem[Zakov {\em et~al.}(2011)Zakov, Tsur, and Ziv-Ukelson]{zakov+:2011}
Zakov, S., Tsur, D., and Ziv-Ukelson, M. (2011).
\newblock Reducing the worst case running times of a family of {RNA} and {CFG}
  problems, using valiant's approach.
\newblock {\em Algorithms for Molecular Biology\/}, {\bf 6}(1), 20.

\bibitem[Zhao {\em et~al.}(2016)Zhao, Li, Fang, Kang, Hao, Li, Bu, Sun, Zhang,
  Chen, {\em et~al.}]{noncode:2016}
Zhao, Y., Li, H., Fang, S., Kang, Y., Hao, Y., Li, Z., Bu, D., Sun, N., Zhang,
  M.~Q., Chen, R., {\em et~al.} (2016).
\newblock Noncode 2016: an informative and valuable data source of long
  non-coding {RNAs}.
\newblock {\em Nucleic Acids Research\/}, {\bf 44}(D1), D203--D208.

\bibitem[Zuker and Stiegler(1981)Zuker and Stiegler]{zuker+stiegler:1981}
Zuker, M. and Stiegler, P. (1981).
\newblock Optimal computer folding of large {RNA} sequences using
  thermodynamics and auxiliary information.
\newblock {\em Nucleic Acids Research\/}, {\bf 9}(1), 133--148.

\bibitem[Zuker {\em et~al.}(1991)Zuker, Jaeger, and Turner]{zuker+:1991}
Zuker, M., Jaeger, J.~A., and Turner, D.~H. (1991).
\newblock A comparison of optimal and suboptimal {RNA} secondary structures
  predicted by free energy minimization with structures determined by
  phylogenetic comparison.
\newblock {\em Nucleic Acids Research\/}, {\bf 19}(10), 2707--2714.

\end{thebibliography}
\onecolumn
\newpage
\setcounter{page}{1}
\renewcommand\thepage{\arabic{page}} 

  \label{sec:supp}

  \begin{centering}
          \textbf{\large Supporting Information}\\[0.2cm]
    \textbf{\large LinearFold: Linear-Time Approximate RNA Folding \\
      by 5'-to-3' Dynamic Programming and Beam Search}\\[0.2cm]
    \textbf{
     \small Liang Huang, He Zhang, Dezhong Deng, Kai Zhao, Kaibo Liu, David Hendrix, and David H.~Mathews}
    
  \end{centering}

\setcounter{figure}{0}
\renewcommand{\thefigure}{SI\,\arabic{figure}} 
\setcounter{table}{0}
\renewcommand{\thetable}{SI\,\arabic{table}}
\setcounter{section}{0}
\renewcommand\thesection{\Alph{section}}

\section{Extra Definitions}
\label{sec:extradefs}

In Section~\ref{sec:formulation}, we sketched the definition of the set of allowed pseudoknot-free secondary structures
\[
\GEN(\vecx) = \big\{\vecy \in \{\md, \ml, \mr\}^{|\vecx|} \mid \balanced(\vecy), \valid(\vecx,\pairs(\vecy))\big\}
\]
Here we complete it.
First we denote $\textstyle\depth(\vecy)=\sum_i \big(\one[y_i=\ml] - \one[y_i=\mr] \big)$ 
to be the difference in counts between ``\ml'' and ``\mr'' in \vecy, and
then $\balanced(\vecy)$ is true iff.:~
\[
\forall i, \depth(y_1 ... y_i) \geq 0; \text{ and } \depth(\vecy) = 0.
\]
We next define the set of pairs in \vecy:
\[
\pairs(\vecy) = \{ (i,j) \mid y_i = \ml, \ y_j = \mr, \ \balanced(y_i ... y_j) \}
\]
and $\valid(\vecx, S)$ checks if all pairs in set $S$ are valid for \vecx, i.e., it returns true iff.:
\[
\forall (i,j) \in S, \ x_i x_j \in \{ \text{CG, GC, AU, UA, GU, UG} \}
\]
We also define $\unpaired(\vecy)=\{ i \mid y_i = \md\}$ to be set of unpaired indices in \vecy. 

\section{Actual Scoring Functions} 

\label{sec:realscore}

The actual scoring functions 
used by \contrafold, \rnafold, and our \linearfold 
decompose into individual loops:

\begin{equation}
  \begin{split}
 \score_{\weight}(\vecx, \vecy) = &
 \!\!\!\!\sum_{(i,j) \in \hairpinloops(\vecy)}\!\!\!\! \score_{\weight}^{\mathrm H}(\vecx, i, j)  + \!\!\!\!\sum_{(i, j, k, l) \in \singleloops(\vecy)}\!\!\!\! \score_{\weight}^{\mathrm S}(\vecx, i, j, k, l) \\
 + & \!\!\!\!\sum_{m \in \multiloops(\vecy)}\!\!\!\! \score_{\weight}^{\mathrm M}(\vecx, m) +  \!\!\!\!\sum_{(i,j) \in \externalloops(\vecy)}\!\!\!\!\!\!\!\! \score_{\weight}^{\mathrm E}(\vecx, i, j).
\end{split}
\label{eq:decomp}
\end{equation}
where $\score_{\weight}^{\mathrm H}(\vecx, \cdot, \cdot)$, $\score_{\weight}^{\mathrm S}(\vecx, \cdot, \cdot, \cdot, \cdot)$, $\score_{\weight}^{\mathrm M}(\vecx, \cdot)$, $\score_{\weight}^{\mathrm E}(\vecx, \cdot, \cdot)$ are scores of hairpin loop, single loop (including bulge and internal loop and stacking), multiloop and external loop, respectively. Multiloop score can be further decomposed into
each adjacent base pair $(i,j)\in m$:
\begin{equation}
  \begin{split}
\score_{\weight}^{\mathrm M}(\vecx, m)=w_{\text {base}}^{\text {multi}} + w_{\text {unpair}}^{\text {multi}} \cdot |\unpaired(m)| + \sum_{(i,j) \in m}{w_{\text {bp}}^{\text {multi}}{(\vecx, i, j)} }
\end{split}
\label{eq:multidecomp}
\end{equation}
For example, if \vecy=\md\ml\md\ml\md\md\md\mr\ml\ml\md\md\md\mr\mr\mr\md, then $\multiloops(\vecy)$ is a singleton-set containing $m=((2,16), (4,8), (9,15))$ with $\unpaired(m)=\{3\}$,
$\hairpinloops(\vecy)=\{(4,8), (10,14)\}$, $\singleloops(\vecy)=\{(9,10,14,15)\}$, and $\externalloops(\vecy)=\{(0,2), (16,17)\}$.

The thermodynamic 
model in \viennarnafold scores 
each type of loop 
using several feature templates such as
hairpin/bulge/internal loop lengths,
terminal mismatches, helix stacking, helix closing, etc.
The machine-learned model in \contrafold
replaces energies in the above framework with model weights learned from data.
Figure~\ref{fig:realdeduct} implement LinearFold for this scoring function.

\section{Extra Results Tables and Figures}

\label{sec:si-tables}

Tables \ref{tab:accuracy} \& \ref{tab:accuracy_nos} detail the accuracy results (PPV \& Sensitivity) from Figure~\ref{fig:accuracy}.
We choose the ArchiveII dataset \cite{sloma+mathews:2016},
a diverse set of over 3,000 RNA sequences with known secondary structures.
But since the current \contrafold machine-learned model (v2.02) is trained on the S-Processed dataset \cite{andronescu+:2007}
we removed those sequences that appeared in the S-Processed dataset. 
The resulting dataset we used contains 2,889 sequences over 9 families, with an average length of 222.2 \nts. 

We sample RNAcentral dataset
by evenly splitting the length range from $1,000$ to
$244,296$ (the longest sequence) into 30 bins by log-scale, and for each bin
randomly select one sequence.


Due to the uncertainty of base-pair matches existing in comparative analysis
and the fact that there is fluctuation in base pairing at equilibrium,
we
consider a base pair to be correctly predicted if it is also displaced by one
nucleotide on a strand \cite{sloma+mathews:2016}.
Generally, if a pair $(i,j)$ is in the predicted structure, we consider it a
correct prediction if one of $(i,j)$, $(i-1,j)$, $(i+1,j)$, $(i,j-1)$, $(i,j+1)$ is in the
ground truth structure.
We also report the accuracy using exact base pair matching instead of this
method, in Table~\ref{tab:accuracy_nos}. 
Both sensitivity and PPV are reported.
Generally, if $\vecyhat$ is the predicted structure and $\vecystar$ is the ground
truth, we have
$\sens = \frac{|\pairs(\vecyhat)\cap\pairs(\vecystar)|}{|\pairs(\vecystar)|}$,
and
$\ppv = \frac{|\pairs(\vecyhat)\cap\pairs(\vecystar)|}{|\pairs(\vecyhat)|}$.

\begin{table*}[h]
  \small
  \centering
  \setlength{\tabcolsep}{4pt}
  \begin{tabular}{r|rrr||rr|rr||rr|rr||rr|rr}
    & \multicolumn{2}{c}{\# of seqs} & avg. & \multicolumn{2}{c|}{\contrafold$^\clubsuit$} & \multicolumn{2}{c||}{\linearfoldc$^\clubsuit$} 
    & \multicolumn{2}{c|}{\contrafold} & \multicolumn{2}{c||}{\linearfoldc}
    & \multicolumn{2}{c|}{\viennarnafold} & \multicolumn{2}{c}{\linearfoldv}\\
    Family & total & used & length & 
    PPV & sens & $\Delta$PPV & $\Delta$sens & 
    PPV & sens & $\Delta$PPV & $\Delta$sens & 
    PPV & sens & $\Delta$PPV & $\Delta$sens \\
    \hline
    tRNA  & 557 & 74  & 77.3  & 68.89 & 70.54 &  +0.00  &  +0.00  & 69.05 & 70.54 & +0.00 & +0.00 & 63.51 & 72.92 &  +0.24  &  +0.19  \\
    5S rRNA & 1,283 & 1,125 & 118.8 & 73.66 & 73.74 &  +0.00  &  +0.00  & 75.52 & 75.61 & +0.00 & +0.00 & 59.55 & 65.96 &  +0.03  &  +0.04  \\
    SRP & 928 & 886 & 186.1 & 62.73 & 62.41 &  -0.07  &  -0.07  & 63.27 & 62.84 & -0.04 & -0.04 & 59.91 & 65.42 &  $^\dagger$+0.35  &  +0.27 \\
    RNaseP  & 454 & 182 & 344.1 & 48.91 & 47.90 & -0.22 &  $^\dagger$-0.54 & 48.96 & 47.67 & -0.11 & -0.14 & 47.28 & 55.15 & +0.12 & -0.07  \\
    tmRNA & 462 & 462 & 366 & 44.88 & 38.61 &  $^\dagger$-0.74 & $^\ddagger$-0.93  & 45.74 & 39.05 &  $^\dagger$-0.67  &  $^\ddagger$-0.82 & 41.47 & 46.86 & $^\ddagger$-0.95  & $^\ddagger$-1.02  \\
    Group I Intron  & 98  & 96  & 424.9 & 52.62 & 50.93 & +0.84 & $^\dagger$+0.80 & 52.36 & 50.64 & +0.87 & +0.80 & 46.81 & 57.68 & $^\ddagger$+0.86  & $^\dagger$+1.02  \\
    telomerase RNA  & 37  & 37  & 444.6 & 45.39 & 59.19 & -0.05 & -0.11 & 45.62 & 59.30 & -0.05 & -0.11 & 41.47 & 58.20 & +0.05 & -0.05\\
    16S rRNA  & 22  & 22  & 1,547.90  & 41.08 & 41.77 & $^\dagger$+3.56 & $^\dagger$+3.09 & 40.20 & 41.21 &  $^\dagger$+3.76  &  $^\dagger$+3.26  & 37.23 & 44.13 & $^\dagger$+1.51 & +1.59 \\
    23S rRNA  & 5 & 5 & 2,927.40  & 52.47 & 53.18 & $^\dagger$+8.65 & $^\dagger$+5.66 & 48.05 & 49.61 &  $^\dagger$+14.03 &  $^\dagger$+9.86  & 54.79 & 62.32 & +0.33 & +0.16 \\
    \hline
    {\em Overall} & 3,846 & 2,889 & 222.2 & 54.51 & 55.36 & +1.33 & +0.88 & 54.31 & 55.16 & +1.98 & +1.42 & 50.22 & 58.74 &  +0.28  & +0.24 \\
  \end{tabular}
  \smallskip
  \caption{Detailed prediction accuracies in percent, allowing one nucleotide in a pair to be displaced by one position, on the ArchiveII dataset using \contrafoldmfe, \linearfoldc, 
    \viennarnafold and \linearfoldv.
    This slipping method~\cite{sloma+mathews:2016}  considers a base pair to
    be correct if it is slipped by one nucleotide on a strand. $^\clubsuit$ denotes using sharpturn enabled mode (default in CONTRAfold).
    Statistical significance are marked by 
    $^\dagger$($0.01\leq p<0.05$) and $^\ddagger$($p<0.01$).
    Overall, \linearfoldc outperforms \contrafoldmfe by +1.33/+0.88 in PPV/sensitivity with sharpturn and by +1.98/+ 1.42 in PPV/sensitivity without sharpturn, 
    and \linearfoldv outperforms \viennarnafold by +0.28/+0.24 in PPV/sensitivity.
    Among the nine families, \linearfoldc is significantly better on three (Group I Intron, 16S and 23S rRNAs), 
    and \linearfoldv is significantly better on three (SRP, Group I Intron, and 16S rRNAs). 
    We also report the accuracies using exact base pair match in the next Table.
    \label{tab:accuracy}}
\end{table*}

\begin{table*}[!h]
  \small
  \centering
  \setlength{\tabcolsep}{4pt}
  \begin{tabular}{r|rrr||rr|rr||rr|rr||rr|rr}
    & \multicolumn{2}{c}{\# of seqs} & avg. & \multicolumn{2}{c|}{\contrafold$^\clubsuit$} & \multicolumn{2}{c||}{\linearfoldc$^\clubsuit$} 
    & \multicolumn{2}{c|}{\contrafold} & \multicolumn{2}{c||}{\linearfoldc}
    & \multicolumn{2}{c|}{\viennarnafold} & \multicolumn{2}{c}{\linearfoldv}\\
    Family & total & used & length & 
    PPV & sens & $\Delta$PPV & $\Delta$sens & 
    PPV & sens & $\Delta$PPV & $\Delta$sens & 
    PPV & sens & $\Delta$PPV & $\Delta$sens \\
    \hline
    tRNA  & 557 & 74  & 77.3  & 67.61 & 69.12 &  +0.00  &  +0.00  & 67.73 & 69.12 & +0.00 & +0.00 & 61.75 & 70.98 &  +0.04  &  -0.07 \\
    5S rRNA & 1,283 & 1,125 & 118.8 & 70.68 & 70.70 &  +0.00  &  +0.00  & 72.60 & 72.59 & +0.00 & +0.00 & 57.28 & 63.35 &  -0.14  &  -0.11 \\
    SRP & 928 & 886 & 186.1 & 59.14 & 58.61 &  -0.05  &  -0.07  & 59.67 & 59.02 & -0.04 & -0.03 & 56.58 & 61.55 &  -0.09  &  -0.20 \\
    RNaseP  & 454 & 182 & 344.1 & 47.45 & 46.39 &  -0.25  &  $^\dagger$-0.55  & 47.49 & 46.15 & -0.13 & -0.15 & 45.76 & 53.28 &  +0.15  &  +0.04 \\
    tmRNA & 462 & 462 & 366 & 42.96 & 36.94 &  $^\dagger$-0.81  &  $^\ddagger$-0.99 & 43.83 & 37.38 &  $^\dagger$-0.72  &  $^\ddagger$-0.85 & 39.75 & 44.90 &  $^\ddagger$-1.09 &  $^\ddagger$-1.17  \\
    Group I Intron  & 98  & 96  & 424.9 & 51.21 & 49.56 &  +0.80  &  $^\dagger$+0.75  & 51.03 & 49.35 & +0.82 & +0.74 & 45.49 & 56.06 &  $^\ddagger$+0.81 & $^\dagger$+0.97  \\
    telomerase RNA  & 37  & 37  & 444.6 & 43.40 & 56.58 &  +0.03  &  +0.00  & 43.66 & 56.72 & +0.04 & +0.00 & 39.53 & 55.40 &  -0.05  &  -0.19 \\
    16S rRNA  & 22  & 22  & 1,547.90  & 39.84 & 40.49 &  $^\dagger$+3.47  &  $^\dagger$+2.99  & 39.01 & 39.97 &  $^\dagger$+3.62  &  $^\dagger$+3.13  & 35.65 & 42.26 &  $^\dagger$+1.33  &  +1.39 \\
    23S rRNA  & 5 & 5 & 2,927.40  & 50.56 & 51.24 &  $^\dagger$+8.51  &  $^\dagger$+5.60  & 46.46 & 47.97 &  $^\dagger$+13.54 &  $^\dagger$+9.47  & 53.20 & 60.50 &  +0.07  &  -0.12  \\
    \hline
    {\em Overall} & 3,846 & 2,889 & 222.2 & 52.54 & 53.29 &  +1.30  & +0.86 & 52.39 & 53.14 & +1.90 & +1.37 & 48.33 & 56.48 & +0.11 & +0.06 \\
  \end{tabular}
  \smallskip
  \caption{The prediction accuracies using exact base-pair matching. 
  Statistical significance are marked by 
    $^\dagger$($0.01\leq p<0.05$) and $^\ddagger$($p<0.01$).
    Overall, \linearfoldc outperforms \contrafoldmfe by +1.30/+0.86 in PPV/sensitivity with sharpturn and by +1.90/+ 1.37 in PPV/sensitivity without sharpturn, 
    and \linearfoldv outperforms \viennarnafold by +0.11 PPV and +0.06 sensitivity. 
    Among the nine families, \linearfoldc is significantly better on three (Group I Intron, 16S and 23S rRNAs), 
    and \linearfoldv is significantly better on two (Group I Intron and 16S rRNAs).  
    \label{tab:accuracy_nos}}
\end{table*}

\begin{figure*}
\centering
\begin{tabular}{ll}
      \panel{A} & \hspace{-.5cm} \panel{B}\\[-0.5cm]
      \includegraphics[width=.26\textwidth]{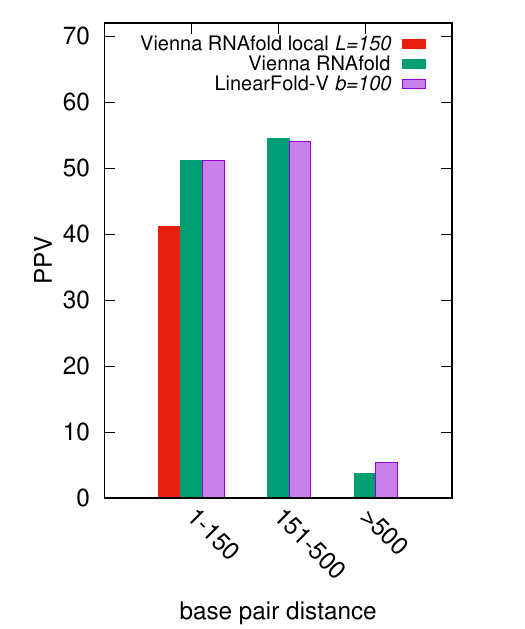}& 
      \hspace{-0.5cm}
      \includegraphics[width=.26\textwidth]{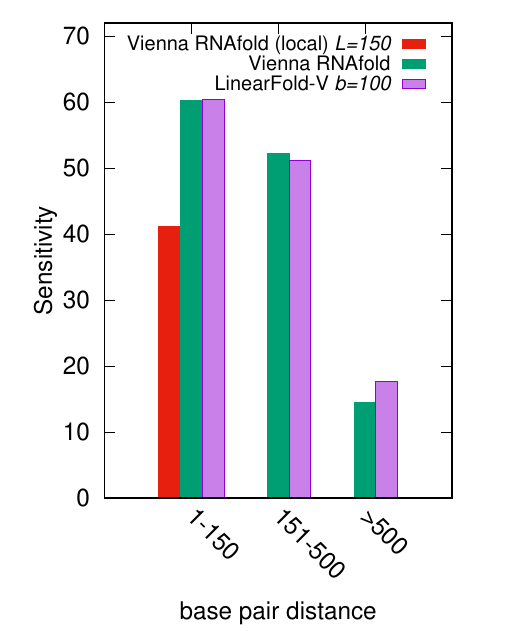} 
\end{tabular}
\vspace{-0.5cm}
      \caption{Comparison of \linearfoldv with \viennarnafold and its local folding mode
      in terms of PPV/Sensitivity of base pairs in certain distance ranges
      across all sequences.
      \linearfoldv is more accurate in long-range base pairs (500+\nts) in both
      PPV and Sensitivity.
      See Fig.~\ref{fig:accuracy}C for the corresponding results for \linearfoldc.
    \label{fig:vienna_distance} }

\end{figure*}

\begin{figure*}
  \centering
   \vspace{-0.2cm}
  \begin{tabular}{lll}
   {\panel{A}} & \hspace{-0.4cm}{\panel{B}} & \hspace{-0.4cm}{\panel{C}}
             \\[-0.2cm]
      \ \ 
      \hspace{-1.5cm}
      \includegraphics[width=.38\textwidth]{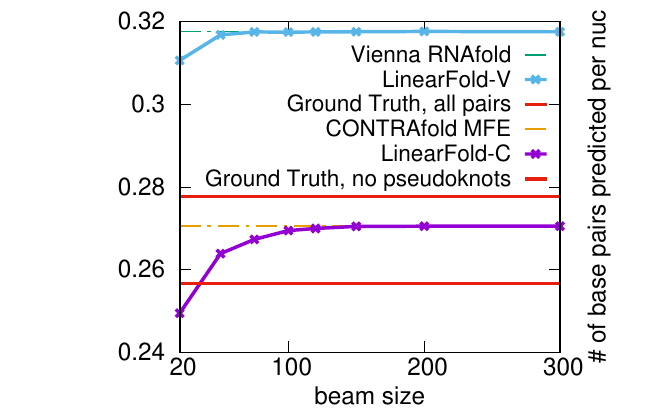}
    &\hspace{-.8cm}
      \includegraphics[width=.35\textwidth]{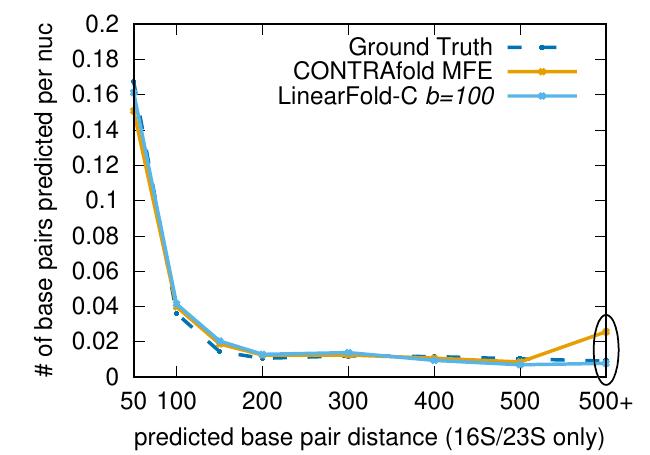}
    &\hspace{-.8cm}
      \includegraphics[width=.35\textwidth]{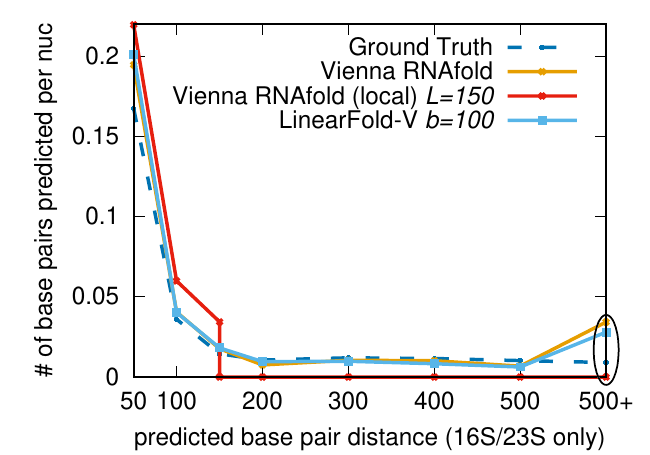}
      \\[-0.4cm]
  \end{tabular}
  \caption{{\bf A}: The number of pairs predicted per nucleotide with varying beam size, 
    comparing these methods and the ground truth (with and without pseudoknots
    (PK)); {\bf B} and {\bf C}: Length distributions of the predicted base pairs using different methods, on the
    16S/23S rRNAs in the ArchiveII dataset.
    Here we plot the number of both predicted and ground truth base pairs (including pseudoknots)
    in each of the following ranges:
    $(0,50], (50,100]$, ... $(400,500)$, $[500,\infty)$.
    This figure shows that \linearfoldc  
    produces almost the same length distributions with the ground truth,
    while \contrafold severely overpredicts base pairs longer than 500\nts apart.
    Both \viennarna and \linearfoldv overpredict in that range, but \linearfoldv is less severe.
    In C, we also reconfirm the limitation of local folding which does not output any long-range pairs.
    \label{fig:bylencnt}}
\end{figure*}

The following Figure details the impact of beam size on the number of pairs predicted.
Figure~\ref{fig:bylencnt}A plots the number of pairs predicted (per nucleotide) with varying beam size,
compared with ground truth (both with and without the pseudoknotted pairs).
It shows that (a) there are on average 0.2776 pairs per nucleotide in this dataset
(meaning about 55.5\% of all nucleotides are paired) and 7.6\% pairs are pseudoknotted;
(b) \viennarna tends to overpredict, while \contrafold tends to underpredict;
(c) our algorithm predicts more pairs with larger beam size;
and (d) with the default beam size, it predicts almost the same amounts of pairs as the baselines
(only 0.0002 and 0.0012 pairs less per nucleotide, respectively).
This is also confirmed by Fig.~\ref{fig:bylencnt}B--C.

\iftrue
\begin{figure*}
  \centering
   \vspace{.5cm}
  \begin{tabular}{ll}
  {\panel{A}} & {\panel{B}}\\[-0.5cm]
      \includegraphics[width=.35\textwidth]{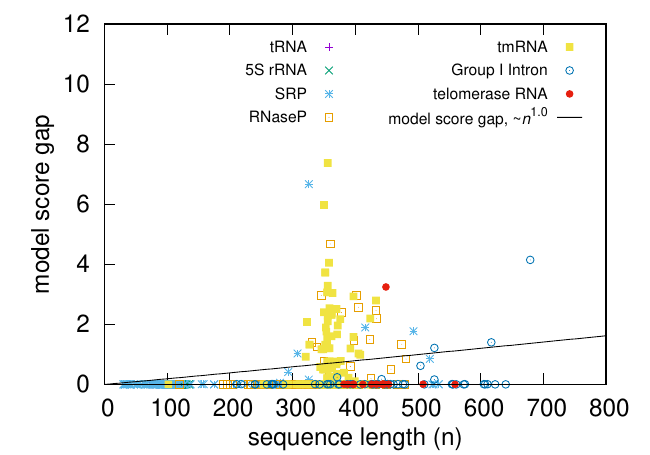} 
    &\hspace{-.2cm}
      \includegraphics[width=.35\textwidth]{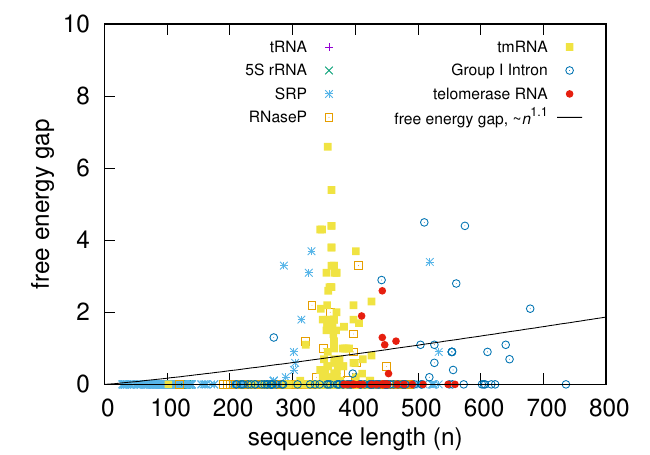} 
      \\[-0.4cm]
  \end{tabular}
  \caption{Close-ups for Fig.~\ref{fig:searcherror} (search error against sequence length)
  for  short sequences.
  A: \linearfoldc vs.~\contrafoldmfe; B: \linearfoldv vs.~\viennarnafold.
  Again, tmRNA is the outlier with disproportionally severe search errors,
  which can explain the slightly worse accuracies of \linearfold on tmRNA
  in Fig.~\ref{fig:accuracy}A. Sequences of 250\nts or less have no search errors
  (i.e., \linearfold with $b$=100 is exact for $n\!\leq\!$ 250).
    \label{fig:searcherror-closeup} }
\end{figure*}
\fi

\iftrue
\begin{figure*}
  \centering
  \vspace{-0.2cm}
  \begin{tabular}{ll}
    {\panel{A}} & {\panel{B}}\\[-0.5cm]
      \includegraphics[width=.35\textwidth]{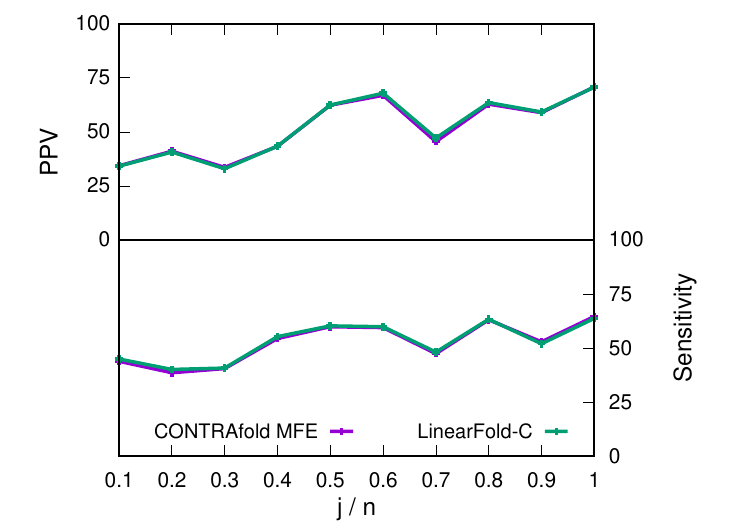} 
    &
      \includegraphics[width=.35\textwidth]{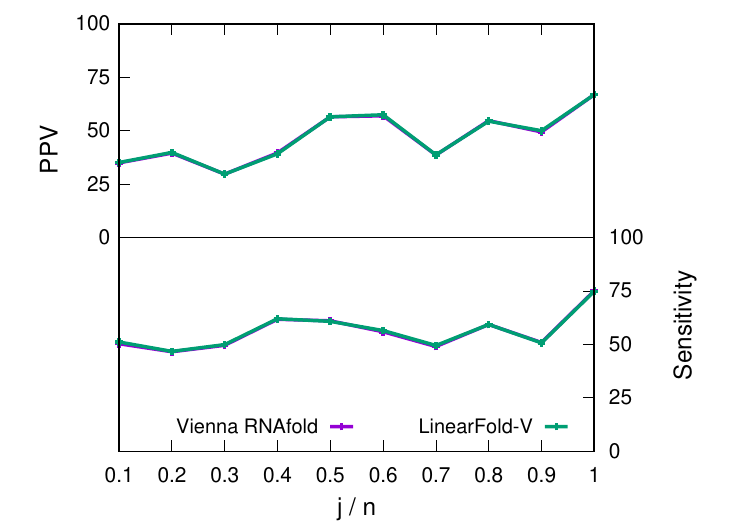} 
      \\[-0.5cm]
  \end{tabular}
  \caption{PPV/Sensitivity for all pairs $(i,j)$ as a function of $j/n$
                           where $n$ is the sequence length,
                           i.e., the ``proportional distance'' of a pair's right nucleotide
                           to the 5'-end.
We bin $j/n$ by $(0,0.1], (0.1,0.2]$,..., $(0.9,1.0]$.
            In general, \linearfold performs very similarly to the baselines,
            and even though it scans 5'-to-3', the accuracy does not degrade towards the 3'-end.
    \label{fig:toward3'} }
\end{figure*}
\fi

\iftrue
\begin{figure}
  \centering
  \begin{tabular}{ll}
  {\panel{A}} & {\panel{B}}\\[-0.5cm]
    \includegraphics[width=.35\textwidth]{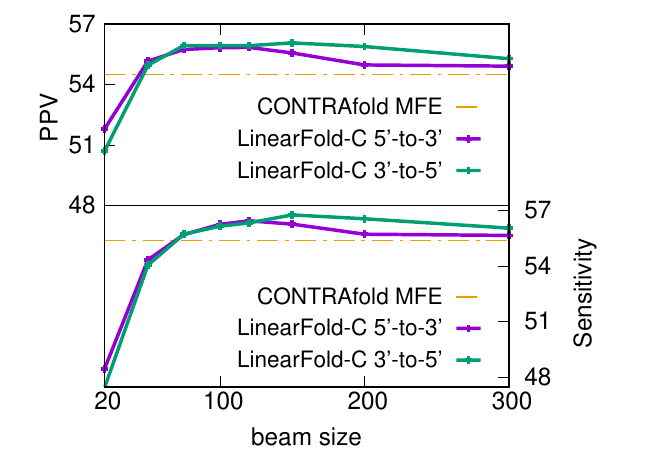} 
    &\hspace{-.2cm}
      \includegraphics[width=.35\textwidth]{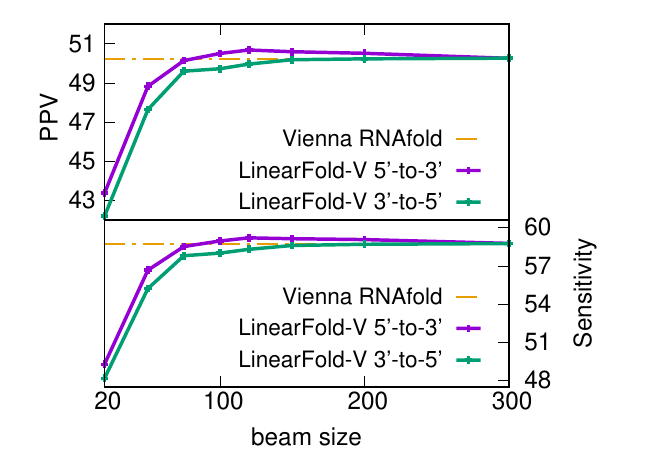} 
      \\[-0.5cm]
  \end{tabular}
  \caption{Comparing 5'-to-3' and 3'-to-5' versions of \linearfold.
  The physical model (B) seems to prefer the default 5'-to-3' order.
    \label{fig:3'-to-5'} }
\end{figure}
\fi

\clearpage

\section{Deductive System for the Actual Systems}

The following Figure sketches the deductive system for the actual LinearFold system based on the real scoring functions in Section~\ref{sec:realscore}.
For more implementation details, we refer the readers to our released source code at
{\myurlsmall{https://github.com/LinearFold/LinearFold}}.

\newcommand{\ShapeE}{\ensuremath{\mathsf{E}}}
\newcommand{\ShapeP}{\ensuremath{\mathsf{P}}}
\newcommand{\ShapeH}{\ensuremath{\mathsf{H}}}
\newcommand{\ShapeM}{\ensuremath{\mathsf{M_1}}}
\newcommand{\ShapeMt}{\ensuremath{\mathsf{M_2}}}
\newcommand{\ShapeMulti}{\ensuremath{\mathsf{M}}}
\newcommand{\hjump}{\ensuremath{\ShapeH\mathsf{jump}}\xspace}
\newcommand{\npair}{\ensuremath{\mathsf{hairpin}}\xspace}
\newcommand{\nreduce}{\ensuremath{\mathsf{reduce}}\xspace}
\newcommand{\single}{\ensuremath{\mathsf{singleloop}}\xspace}
\newcommand{\multi}{\ensuremath{\mathsf{multiloop}}\xspace}
\newcommand{\multileft}{\ensuremath{\ShapeMulti\mathsf{left}}\xspace}
\newcommand{\multijump}{\ensuremath{\ShapeMulti\mathsf{jump}}\xspace}

\newcommand{\xtom}{\ensuremath{\mathsf{XtoM_1}}\xspace}
\newcommand{\myus}{\ensuremath{\text{\tt \_\,}}\xspace}

\begin{figure}[!h]
  \centering
  \vspace{-0.3cm}
  \scalebox{1.1}{
  \hspace {-1.3em}
    \begin{tabular}{lll}
      input      & $x_1 \ldots x_n$ &  \\[0.1in]
      states & $\ShapeE$ \nitemt{0}{j}{\codeblue{$\alpha$}, s}  &  prefix structure \\[0.1in] 
                 & $\ShapeP$ \nitemt{i}{j}{\codeblue{\ml$\alpha$\mr}, s} & pair \\[0.1in] 
                 & $\ShapeH$ \nitemt{i}{j}{\bml\codeblue{\md\md\md},s} & hairpin candidate \\[0.1in] 
                 & $\ShapeM$ \nitemt{i}{j}{\codeblue{\ml$\alpha$\mr$\beta$},s} & one or more pairs \\[0.1in]
                 & $\ShapeMt$ \nitemt{i}{j}{\codeblue{\ml$\alpha$\mr$\beta$\ml$\gamma$\mr},s} & two or more pairs \\[0.1in] 
                 & $\ShapeMulti$ \nitemt{i}{j}{\bml\codeblue{\md\md\md\ml$\alpha$\mr$\beta$\ml$\gamma$\mr\md\md\md},s} &  multiloop candidate \\[0.2in] 
                 
      axiom & $\ShapeE$ \nitemt{0}{1}{\codeblue{\!},0} 
      & goal \qquad $\ShapeE$ \nitemt{0}{n\!\!+\!1}{\codeblue{$\alpha$},\_} \\[0.1in]
      {\push}  & \inferrule{\ \ \ \ \ \ \ \ \ \  \ShapeE \nitemt{0}{j}{\codeblue{$\alpha$},s}}{ \ShapeH \nitemt{j}{\jnext(j,j)}{\bml\codeblue{\md\md}, 0}} & $\jnext(i,j) \triangleq \min\{k \mid k>j, \ (x_i,x_{k})~\text{match}\}$ \\[0.2in] %
      {\hjump} & \inferrule{\ \ \ \ \ \ \ \ \ \ \ShapeH \nitemt{i}{j}{\bml\codeblue{\md\md\md}, s}}{ \ShapeH \nitemt{i}{\jnext(i,j)}{\bml\codeblue{\md\md\md\md\md}, s}} \\[0.2in]
      {\nskip}   & \inferrule{ \ShapeE \nitemt{0}{j}{\codeblue{$\alpha$}, s} \ \ \ \ \ \ }{ \ShapeE \nitemt{0}{j\!\!+\!1}{\codeblue{$\alpha$\md}, s \!\!+\! \score_\weight^{\mathrm {E}}{(\vecx, j, j+1) }}} 
                 & \inferrule{ \ShapeM \nitemt{i}{j}{\codeblue{$\ml\alpha\mr\beta$}, s} \ \ \ \ \ \ }{ \ShapeM \nitemt{i}{j\!\!+\!1}{\codeblue{\ml$\alpha\mr\beta$\md}, s \!\!+\! w_{\text {unpair}}^{\text {multi}}  }} \\[0.2in]
      {\nreduce} & \inferrule{ \ShapeM \nitemt{k}{i}{\codeblue{$\ml\alpha\mr\beta$}, s'} \quad \quad \ShapeP \nitemt{i}{j}{\codeblue{$\ml\gamma\mr$}, s}}{ \ShapeMt \nitemt{k}{j}{\codeblue{$\ml\alpha\mr\beta\ml\gamma\mr$} , s'\!\!+\!s\!\!+\!w_{\text {bp}}^{\text {multi}}(\vecx, i, j) }} \\[0.2in]

      {\ncombine} & \inferrule{ \ShapeE \nitemt{0}{i}{\codeblue{$\alpha$},s'} \quad\quad \ShapeP \nitemt{i}{j}{\codeblue{$\ml\beta\mr$}, s}}{ \ShapeE \nitemt{0}{j}{\codeblue{$\alpha\ml\beta\mr$}, s'\!\!+\!s\!\!+\!{\rm {sc}}_{\weight}^{\mathrm E}{({\vecx}, i, j)} }} \\[0.2in]
      {\xtom} & \inferrule { \ShapeP \nitemt{i}{j}{\codeblue{$\ml\alpha\mr$},s} \ \ \ \ \ \ }{ \ShapeM \nitemt{i}{j}{\codeblue{$\ml\alpha\mr$},s\!\!+\!w_{\text {bp}}^{\text {multi}}(\vecx, i, j)}} 
      & \inferrule { \ShapeMt \nitemt{i}{j}{\codeblue{$\ml\alpha\mr\beta\ml\gamma\mr$},s}}{ \ShapeM \nitemt{i}{j}{\codeblue{$\ml\alpha\mr\beta\ml\gamma\mr$},s}} \\[0.2in]
      {\multileft} & \inferrule{\ShapeMt \nitemt{i}{j}{\codeblue{$\ml\alpha\mr\beta\ml\gamma\mr$}, s} \ \ \ \ \ \ \ \ \ \ \ \ \ }{\ShapeMulti \nitemt{k}{\jnext(k,j)}{ \bml\codeblue{$\md\md\md\ml\alpha\mr\beta\ml\gamma\mr\md\md$}, s\!\!+\!u\cdot w_{\text {unpair}}^{\text {multi}} 
      } } &  $u = (\jnext(k,j)\!-\!j)\!\!+\!(i\!-\!k\!-\!1)$, \\&&$i\!-\!k\!-\!1\leq 30$\\[0.2in]
      {\multijump} & \inferrule{\ShapeMulti \nitemt{i}{j}{\bml\codeblue{$\md\md\md\ml\alpha\mr\beta\ml\gamma\mr\md\md\md$}, s} \ \ \ \ }{\ShapeMulti \nitemt{i}{\jnext(i,j)}{\bml\codeblue{$\md\md\md\ml\alpha\mr\beta\ml\gamma\mr\md\md\md\md\md$}, s\!\!+\! \!u\cdot w_{\text {unpair}}^{\text {multi}} 
      }  } & $u = \jnext(i,j) \!\!-\! j$ \\[0.2in]

      {\npair} & \inferrule{ \ShapeH \nitemt{i}{j}{\bml\codeblue{$\md\md\md$},s} \ \ }{ \ShapeP \nitemt{i}{j\!\!+\!1}{\codeblue{\ml\md\md\md\mr}, s\!\!+\!\score_{\weight}^{\mathrm H}(\vecx, i, j)}} \\[0.2in]

      {\single} & \inferrule{ \ShapeP \nitemt{i}{j}{\codeblue{$\ml\alpha\mr$},s} \ \ \ \ \ \ \ \ \ }{ \ShapeP \nitemt{k}{l}{ \codeblue{\ml\md\md\md$\ml\alpha\mr$\md\md\md\mr},s\!\!+\!{\score_{\weight}^{\mathrm S}(\vecx, i, j, k, l)}}} &  $(x_k, x_{l\!-\!1})~$ match, $(l\!-\!j\!-\!1)\!\!+\!(i\!-\!k\!-\!1) \leq 30$ \\[0.2in]

      {\multi} & \inferrule{ \ShapeMulti \nitemt{i}{j}{\bml\codeblue{$\md\md\md\ml\alpha\mr\beta\ml\gamma\mr\md\md\md$},s}\ \ \ \ \ \ }{ \ShapeP \nitemt{i}{j\!\!+\!1}{\codeblue{$\ml\md\md\md\ml\alpha\mr\beta\ml\gamma\mr\md\md\md\mr$}, s\!\!+\! w_{\text {base}}^{\text {multi}} \!+\! w_{\text {bp}}^{\text {multi}}(\vecx, i, j) }} 
      \\[-0.1in]
    \end{tabular}
  }
  \caption {The actual deductive system implemented in \linearfold.
  Shaded substrings are balanced in brackets.
  Here $\score_\weight^{\mathrm E}(\vecx, \cdot, \cdot)$, $w_{\text {base}}^{\text {multi}}$, $w_{\text {bp}}^{\text {multi}}{(\vecx, \cdot, \cdot)}$, $w_{\text {unpair}}^{\text {multi}}$, $\score_\weight^{\mathrm S}(\vecx, \cdot, \cdot, \cdot, \cdot)$, $\score_\weight^{\mathrm H}(\vecx, \cdot, \cdot)$ are the various energy or scoring parameters (E stands for external loop, multi for multiloop, S for single loop, and H for hairpin loop).
  The $\jnext(i, j)$ returns the next position after $x_j$ that can pair with $x_i$;
  this is the ``jumping'' trick used in \contrafold and \viennarna.
  Our final two rules also use this jumping trick in the righthand side loop.
  The only cubic-time rule is \nreduce (intermediate step in multiloop), again inspired by \contrafold source code.
    \label{fig:realdeduct}}
\end{figure}


\clearpage

\section{Connections between Context-Free Parsing and RNA Folding}

\definecolor {processblue}{cmyk}{0.96,0,0,0}
\begin{figure}[h]
  \centering
  \resizebox{0.9\textwidth}{!}{
	\begin {tikzpicture}[-latex ,auto ,node distance =4 cm and 5cm ,on grid ,
		semithick ,
		state/.style ={ rectangle ,top color =white , bottom color = processblue!20 ,
		draw,processblue , text=blue , minimum width =3 cm}
		tcancel/.append style={rectangle, processblue, draw=#1, cross out, inner sep=1pt}]
	  \node[state] (cky) {\begin{tabular}{c}
              CKY parsing: $O(n^3)$\\
              \citet{kasami:1965}\\
              \citet{younger:1967}
          \end{tabular}};

          \node[state] (NLP) [above=1cm of cky] {\bf natural language parsing};
          
	  \node[state] (knuth) [left=5cm of cky] {\begin{tabular}{c}
              \em LR parsing: $O(n)$\\
              \em \citet{knuth:1965}
          \end{tabular}};

          \node[state] (PL) [above=1cm of knuth] {\bf programming language parsing};

	  \node[state] (tomita) [below=1.5cm of cky] {\begin{tabular}{c}
              \em Generalized LR: $O(n^3)$\\
              \em \citet{tomita:1988}
              \end{tabular}
          };
	  \node[state] (hns) [below=1.5cm of tomita] {\begin{tabular}{c}
              \em Approximate DP: $O(n)$\\
              \em \citet{huang+sagae:2010}
              \end{tabular}
          };

	\node[state] (nussinov) [right=5cm of cky]{\begin{tabular}{c}
            classical RNA folding: $O(n^3)$\\
            \citet{nussinov+:1978}\\
            \citet{zuker+stiegler:1981}\end{tabular}};

        \node[state] (rna) [above=1cm of nussinov] {\bf RNA folding};

	\node[state] (idea2) [right=5cm of tomita]{\begin{tabular}{c}
            \em LinearFold: Idea 2\\
            \em exact $O(n^3)$
            \end{tabular}
        };
	\node[state] (linearfold) [right=5cm of hns]{\begin{tabular}{c}
            \em LinearFold: Idea 3\\
            \em approx.~$O(n)$ 
            \end{tabular}
        };

	\draw[=>, thick, blue] (cky) -- node [fill=white, inner sep=1pt, midway] {} (tomita); 
	\draw[=>, thick, blue] (knuth) -- node [fill=white, inner sep=1pt, midway] {} (tomita); 
	\draw[=>, thick, blue] (tomita) -- node [fill=white, inner sep=1pt, midway] {} (hns); 

	\draw[=>, thick, blue] (nussinov) -- node [fill=white, inner sep=1pt, midway] {} (idea2); 
	\draw[=>, thick, blue] (idea2) -- node [fill=white, inner sep=1pt, midway] {} (linearfold); 

	\draw[-, thick, dashed, blue] (cky) -- node [fill=white, inner sep=1pt, midway] {} (nussinov); 
	\draw[=>, thick, blue] (tomita) -- node [fill=white, inner sep=1pt, midway] {} (idea2);
	\draw[=>, thick, blue] (hns) -- node [fill=white, inner sep=1pt, midway] {} (linearfold); 
	\end{tikzpicture}
        }
  \caption{Our work 
    is inspired by
    incremental parsing algorithms in both programming language theory
    and computational linguistics. 
     Left-to-right algorithms are in {\em italic}; others are bottom-up.
    The classical bottom-up $O(n^3)$ algorithms are isomorphic between natural language parsing
    and RNA folding.
    Knuth's $O(n)$ LR algorithm works only for a small subset of context-free grammars (CFGs),
    and Tomita generalizes it to arbitrary CFGs, achieving the alternative, left-to-right,
    $O(n^3)$ algorithm, which inspires \linearfold \hyperref[sec:idea2]{Idea 2}.
    Our previous work (Huang and Sagae) modernize and generalize
    Tomita's algorithm, combining it with beam search to achieve
    linear runtime, which inspires \linearfold \hyperref[sec:idea3]{Idea 3}.
    \label{fig:connect}
  	}

\end{figure}
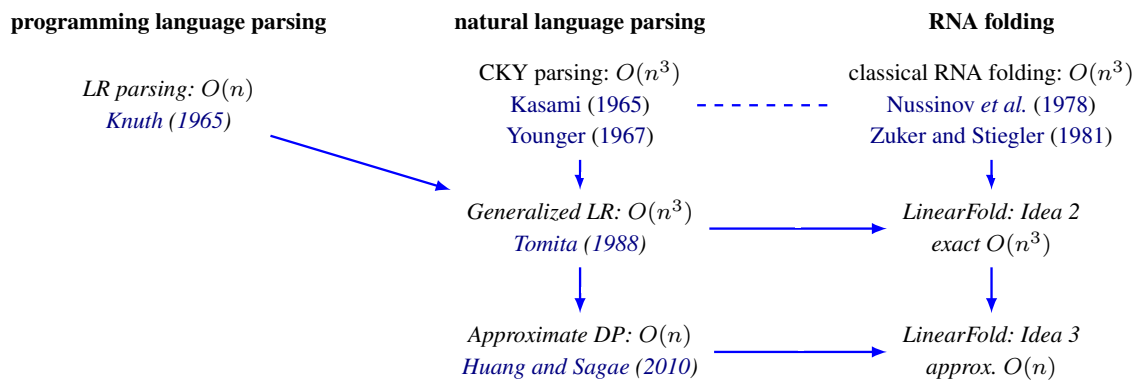

\end{document}